\pdfoutput=1
\documentclass[preprint, 5p, times]{elsarticle}
\usepackage{graphicx}
\usepackage{amsmath, amssymb}

\usepackage[utf8]{inputenc}
\usepackage[T1]{fontenc}

\usepackage[protrusion=true,expansion=true]{microtype} 

\usepackage[pdftex, bookmarks = true]{hyperref}
\hypersetup{
pdftitle = {First-order frequency-shifts in a Penning Trap}, 
pdfauthor = {Jochen Ketter et al.}, 
pdfsubject = {Accepted manuscript}, 
pdfkeywords = {Penning trap; mass spectrometry; perturbation theory; electrostatics; magnetostatics}, 
pdfstartview = {FitH},
}

\biboptions{sort&compress}

\makeatletter
\def\ps@pprintTitle{%
 \let\@oddhead\@empty
 \let\@evenhead\@empty
 \def\@oddfoot{\footnotesize\itshape 
       Revised manuscript subsequently published as \href{http://dx.doi.org/10.1016/j.ijms.2013.10.005}{DOI:10.1016/j.ijms.2013.10.005} \hfill \today}%
 \let\@evenfoot\@oddfoot}
\makeatother

\newcommand*{\myd}{\mathrm{d}} 
\newcommand*{\myi}{\mathrm{i}} 
\newcommand*{\myGSAbstand}{\ } 
\newcommand*{\myank}[2]{a_{#1}(#2)} 
\newcommand*{\myatnk}[2]{\tilde{a}_{#1}(#2)}  
\newcommand*{\myAmpR}[1]{\hat{\rho}_{#1}} 
\newcommand*{\myAmpZ}{\hat{z}} 
\newcommand*{\myvec}[1]{\vec{#1}} 
\newcommand*{\mybeat}{\mathrm{b}} 
\newcommand*{\myeuv}[1]{\vec{e}_{#1}} 
\newcommand*{\myomc}{\omega_\mathrm{c}} 
\newcommand*{\myomz}{\omega_z} 
\newcommand*{\myPhiT}[1]{\omega_{#1}t+\varphi_{#1}} 
\newcommand*{\myPhiTot}[1]{\chi_{#1}} 
\newcommand*{\myomzw}{\tilde{\omega}_z} 
\newcommand*{\myomm}{\omega_-} 
\newcommand*{\myommw}{\tilde{\omega}_-} 
\newcommand*{\myomp}{\omega_+} 
\newcommand*{\myompw}{\tilde{\omega}_+} 
\newcommand*{\myompm}{\omega_\pm} 
\newcommand*{\myommp}{\omega_\mp} 
\newcommand*{\myomw}{\tilde{\omega}} 
\newcommand*{\myompmw}{\tilde{\omega}_\pm} 
\newcommand*{\myommpw}{\tilde{\omega}_\mp} 
\newcommand*{\myombw}{\tilde{\omega}_\mathrm{b}} 
\newcommand*{\mynabla}{\vec{\nabla}} 
\newcommand*{\myfComp}[2]{\left\langle {#1} \right\rangle_{#2}} 
\newcommand*{\myfZero}{0} 
\newcommand*{\myPhiTw}[1]{\tilde{\omega}_{#1}t+\tilde{\varphi}_{#1}} 
\newcommand*{\myinn}{\eta} 
\newcommand*{\myPhiTotw}[1]{\tilde{\chi}_{#1}} 
\newcommand*{\myCnfactor}[1]{\kappa_{#1}} 
\newcommand*{\myZOT}[1]{\tilde{#1}} 
\newcommand*{\myhot}{\text{set}}
\newcommand*{\mycool}{\text{cool}}
\newcommand*{\mySec}[1]{Section~\ref{#1}}
\newcommand*{\mySecs}[1]{Sections~\ref{#1}}
\newcommand*{\mySubsec}[1]{Subsection~\ref{#1}}
\newcommand*{\myEqua}[1]{Equation~\eqref{#1}}
\newcommand*{\myEquas}[1]{Equations~\eqref{#1}}
\newcommand*{\myFig}[1]{Figure~\ref{#1}}

\begin{document}
\begin{frontmatter}

\title{First-order perturbative calculation of the frequency-shifts caused by static cylindrically-symmetric electric and magnetic imperfections of a Penning trap}

\author{Jochen Ketter\corref{cor1}}
\ead{jochen.ketter@mpi-hd.mpg.de}
\cortext[cor1]{Corresponding author}
\author{Tommi Eronen\corref{cor2}}
\author{Martin Höcker}
\author{Sebastian Streubel}
\author{Klaus Blaum}

\address{Max-Planck-Institut für Kernphysik, Saupfercheckweg 1, 69117 Heidelberg, Germany}

\begin{abstract}
The ideal Penning trap consists of a uniform magnetic field and an electrostatic quadrupole potential. 
Cylindrically-symmetric deviations thereof are parametrized by the coefficients~$B_{\eta}$ and $C_{\eta}$, respectively. 
Relativistic mass-increase aside, the three characteristic eigenfrequencies of a charged particle stored in an ideal Penning trap are independent of the three motional amplitudes. 
This threefold harmonicity is a highly-coveted virtue for precision experiments that rely on the measurement of at least one eigenfrequency in order to determine fundamental properties of the stored particle, such as its mass. 
However, higher-order contributions to the ideal fields result in amplitude-dependent frequency-shifts. 
In turn, these frequency-shifts need to be understood for estimating systematic experimental errors, and eventually for correcting them by means of calibrating the imperfections. 
The problem of calculating the frequency-shifts caused by small imperfections of a near-ideal trap yields nicely to perturbation theory, producing analytic formulas that are easy to evaluate for the relevant parameters of an experiment. 
In particular, the frequency-shifts can be understood on physical rather than purely mathematical grounds by considering which terms actually drive them. 
Based on identifying these terms, we derive general formulas for the first-order frequency-shifts caused by any perturbation parameter~$B_{\eta}$ or $C_{\eta}$.
\end{abstract}



\end{frontmatter}

\section{Introduction} 
Much more than a device for storing charged particles~\cite{blaum2010}, the Penning trap excels at relating fundamental properties of the stored particle, such as its mass or magnetic moment, to a measurable frequency~\cite{blaum2006}. 
In order to make full use of the precision the Penning trap has to offer, the relationship between the measured frequency and the sought-for quantity has to be understood in detail despite the complications that come with a real-world experiment. 
Deviations from the ideal Penning trap may be unavoidable in general, but they can also serve a purpose as a part of the detection system~\cite{dehmelt1986, moore1992}. 

In this paper, we employ a perturbative method to deal with one particularly important subset of imperfections---cylindrically-symmetric ones---and we focus on the frequency-shifts they cause. 
Although these are not the only consequence of imperfections, the frequency-shifts are often the most significant one, considering that frequencies constitute the main observables in a typical experiment. 

Although the frequency-shifts for the experimentally most relevant lowest-order cylindrically-symmetric imperfections have previously been given numerous times~\cite{brown1986gtp,bollen1990,gerz1990,mitchell1995ele,mitchell1995mag}, and the prescriptions for calculating all the first-order shifts caused by this subset of imperfections have been outlined in general~\cite{kretzschmar1990, kretzschmar1992, thompson2003, rainville2003}, the specific formulas lack the common ground a general expression would provide. 
In this paper, we derive such readily-evaluated general expressions for all the first-order frequency-shifts caused by cylindrically-symmetric imperfections.  
As a little known fact, a general treatment of the problem has been attempted before~\cite{cptI2005} with Hamiltonian perturbation theory and classical canonical action-variables, but since we disagree with the result given for magnetic imperfections, a complete and correct check is certainly welcome. 
Moreover, we try to be more explicit about our calculation, thereby allowing the reader to verify its validity. 

In~\mySec{sec:IdealPT}, we review the most important properties of the ideal Penning trap as the zeroth-order input for the perturbative treatment of imperfections. 
\mySec{sec:ParaImp} then deals with how to parametrize cylindrically-symmetric electric and magnetic imperfections. 
The mathematical groundwork for the calculation is laid in \mySec{sec:TheoFrame} with particular emphasis on the implementation of perturbation theory. 
With this method outlined in \mySubsec{subsec:PertTheory}, the actual first-order frequency-shifts are subsequently calculated in \mySec{sec:FreqShiftC} and \mySec{sec:FreqShiftB} for electric and magnetic imperfections, respectively. 

\section{The ideal Penning trap} \label{sec:IdealPT}
The ideal Penning trap consists of a homogeneous magnetic field $\myvec{B}_0 = B_0\myeuv{z}$ pointing along the $z$-axis and an electrostatic quadrupole potential
\begin{gather}
\Phi_2(\rho, z) = \frac{V_0C_2}{2d^2}\left(z^2 -\frac{\rho^2}{2}\right) \myGSAbstand\text{, where } \rho = \sqrt{x^2+y^2} \label{eq:IPT-quadPot}
\end{gather}
is the distance from the $z$-axis. 
In the context of the experiment, $V_0$ is understood as an applied voltage. 
The characteristic trap dimension~$d$ is typically defined such that the dimensionless parameter~$C_2$ is close to unity for traps with hyperboloidal electrodes~\cite{gabrielse1983rce}, but any value may be used to describe the quadrupole contribution in other trap geometries, such as cylindrical traps with flat-plate~\cite{gabrielse1984cpt} or open endcaps~\cite{gabrielse1989}. 

Throughout this paper, we will work with the classical Newtonian equation of motion 
\begin{gather}
\ddot{\myvec{r}} = \frac{\vec{F}_\mathrm{L}}{m} = \frac{q}{m}\left(\myvec{E}+\dot{\myvec{r}}\times \myvec{B} \right) \label{eq:CNewton-EQM}
\end{gather}
into which we insert the Lorentz force~$\vec{F}_\mathrm{L}$ experienced by a point-like particle of mass~$m$ and charge~$q$ in the magnetic field~$\myvec{B}$ and the electric field $\myvec{E}= - \mynabla \Phi$, derived by taking the negative gradient of the electrostatic potential~$\Phi$. 
Since the equation is linear in both fields, we will simply add imperfections as we go along. 

While early treatment of the ideal Penning trap partly started out from a quantum-mechanical perspective~\cite{graeff1967, sokolov1967}, and an operator formalism suits the excitation and coupling of modes well~\cite{kretzschmar2007, kretzschmar2011}, we will content ourselves with a purely classical model, ignoring both quantum-mechanical and relativistic effects. 
Spin and relativistic mass-increase can be treated as a perturbation of their own~\cite{brown1986gtp}. 
Furthermore, we will restrict ourselves to  the ``static'' case, meaning that the particle oscillates with constant motional amplitudes in the absence of external excitation drives. 

The emission of synchrotron radiation by an electron orbiting in a strong magnet field allows to cool the electron's cyclotron motion into its quantum-mechanical ground-state. 
For heavier particles, radiative cooling is inefficient~\cite{brown1986gtp}, and the motional ground-state remains out of reach unless laser-cooling is used~\cite{wineland1978}. 
Typical other techniques such as buffer-gas cooling~\cite{savard1991}, resistive cooling of one motion via an $LC$ tank circuit~\cite{dehmelt1968}, and cooling via sideband-coupling to a cooled motion~\cite{dehmelt1976} leave the particle with high enough a set of quantum numbers to warrant a classical treatment. 
Moreover, some detection methods rely on motional amplitudes well above the thermal limit. 
It is only recently that quantum-jumps in the motion of a single resistively-cooled proton are on the brink of being resolved in a huge magnetic inhomogeneity, albeit as a spurious and ill-controlled side-effect where spin-flips are to be detected~\cite{mooser2013, diSciacca2013}. 

Throughout this paper, we will assume a charged particle devoid of internal degrees of freedom which could couple to electric or magnetic fields. 
Apart from spin, this also excludes polarizability, which may play the role of an effective mass~\cite{thompson2004}. 
For the ideal Penning trap with $\myvec{B}_0 = B_0\myeuv{z}$ and $\myvec{E}_2= -\mynabla \Phi_2$, the classical equations of motion for a particle of charge~$q$ and mass~$m$ are
\begin{align}
\begin{pmatrix}\ddot{x}\\\ddot{y}\\\ddot{z}\end{pmatrix} 
&=\frac{qB_0}{m} \begin{pmatrix}\dot{y}\\-\dot{x}\\0\end{pmatrix} + \frac{qV_0C_2}{2md^2}\begin{pmatrix}x\\y\\-2z\end{pmatrix} \label{eq:EQM-IPT} \myGSAbstand .
\end{align}
Being parallel to and therefore unaffected by the magnetic field, the axial motion is a one-dimensional harmonic oscillator with the angular frequency
\begin{gather}
\myomz = \sqrt{\frac{qV_0C_2}{md^2}} \label{eq:uptFreqAxial}\myGSAbstand .
\end{gather}
Trapping requires $qV_0 C_2 > 0$. 
If there was no electric field, the particle would orbit around the magnetic field-lines with the free-space cyclotron-frequency
\begin{gather}
\myomc = \frac{qB_0}{m} \label{eq:fscycf} \myGSAbstand .
\end{gather}
For $V_0 \ne 0$, the radial motion consists of two circular modes with frequencies%
\footnote{%
In contrast to virtually all other publications, we have included essentially the sign of $\myomc$ as a prefactor of the square root in \myEqua{eq:uptRadFreq}, which allows us to handle negative cyclotron frequencies consistently. 
Whereas the sign of the angular frequency is not an additional degree of freedom for the one-dimensional axial motion and was consequently taken to be positive by convention, the sign of the angular frequencies associated with the two-dimensional radial motions encodes the sense of revolution in a natural manner. 
We do not have to think about the sign of the charge $q$ or of the magnetic field $B_0$, which could point along the negative $z$-axis. 
Therefore, we will not work with true frequencies $\nu = \frac{\lvert\omega\vert}{2\pi}$ in this paper. 
This is not meant to imply that the sense of rotation, defined in a coordinate system with either of two possible choices for the $z$-axis, impacts the frequency-shift---it does not, even less so for cylindrically-symmetric imperfections. 
However, the sign of the perturbation parameter $B_{\myinn}$ with respect to $B_0$ matters, and we do not want to run the risk of losing it while working with the absolute value of $qB_0$ in the free-space cyclotron-frequency $\nu_\mathrm{c}$. 
Moreover, the definition of $\omega_\pm$ with the additional factor $\myomc/\lvert\myomc\rvert$ ensures that $\lvert\myomp\rvert \ge \lvert\myomm\rvert$, regardless of the sign of $\myomc$.%
}  
\begin{align}
\omega_\pm &= \frac{1}{2}\left(\myomc \pm \frac{\myomc}{\lvert\myomc\rvert}\sqrt{\myomc^2-2\myomz^2}\right) \label{eq:uptRadFreq} \myGSAbstand .
\end{align}
Because the frequencies have to be real for the motion to stay bounded, the second condition for trapping is $\vert \myomc\vert > \sqrt{2} \myomz$. 
The radial mode with the lower (absolute) frequency is called magnetron motion; 
the frequency~$\myomp$ is associated with the modified cyclotron motion and also referred to as the reduced cyclotron-frequency because its absolute value is lower than the free-space cyclotron-frequency~$\myomc$. 
In a typical experiment, the hierarchy is $\lvert\myomc\rvert \gtrsim \lvert\myomp\rvert \gg \myomz \gg \lvert\myomm\rvert$. 

The trajectory in the ideal Penning trap is given by 
\begin{align}
x(t) &= \myAmpR{+} \cos(\myPhiT{+}) + \myAmpR{-} \cos(\myPhiT{-}) \label{eq:x0} \myGSAbstand, \\
y(t) &= -\myAmpR{+} \sin(\myPhiT{+}) - \myAmpR{-} \sin(\myPhiT{-})  \label{eq:y0} \myGSAbstand,\\
z(t) &= \myAmpZ\cos(\myPhiT{z})  \label{eq:z0} \myGSAbstand.
\end{align}
The amplitudes~$\myAmpR{\pm}$ of the two radial modes and the amplitude~$\myAmpZ$ of the axial mode, as well as the corresponding initial phases~$\varphi_i$ with $i=(+,-,z)$ are determined by the initial conditions. 
Later on, we will use
\begin{gather}
\myPhiTot{i} = \myPhiT{i} \label{eq:xit} 
\end{gather}
as an abbreviation for the total phase without always stressing the time-dependent nature of~$\myPhiTot{i}$.

From \myEqua{eq:uptRadFreq}, we derive the three relations 
\begin{align}
\myomp + \myomm &= \myomc \label{eq:SB-Cyc}\myGSAbstand ,\\
2\myomp  \myomm &= \myomz^2 \label{eq:omzq-omrad} \myGSAbstand ,\\
\myomp^2+\myomm^2+\myomz^2 &= \myomc^2 \label{eq:invt} \myGSAbstand.
\end{align}
The first and the last identity are particularly important, not only because they relate the  eigenfrequencies in the Penning trap to the free-space cyclotron-frequency~$\myomc$. 
Given the extraordinary stability achievable with superconducting magnets~\cite{gabrielse1988, vanDyck1999}, the magnetic field is the gold standard for connecting the eigenfrequencies of a particle to its mass. 

Note that the ideal Penning trap is inherently harmonic. 
Relativistic mass-increase aside, none of the three eigenfrequencies depends on the amplitudes of any eigenmotion. 
This is different for the anharmonic shifts we will deal with in this paper, all of which have a distinct dependence on the particle's motional amplitudes. 
Apart from these higher-order contributions, the magnetic field in a Penning trap may be misaligned with respect to the electrostatic potential, which may also have an elliptic contribution of the kind $(x^2-y^2)$~\cite{kretzschmar2008}. 
These two imperfections lead to frequency-shifts that do not depend on the particle's motional amplitudes. 
Although \myEqua{eq:SB-Cyc} for the so-called sideband cyclotron-frequency is only valid in the ideal Penning trap, the effects of misalignment and ellipticity are suppressed to such an extent that anharmonic effects become dominant~\cite{gabrielse2009}.  
\myEqua{eq:invt}, the so-called Brown--Gabrielse invariance theorem~\cite{brown1982psc}, holds despite misalignment and ellipticity, but it offers no cure for anharmonic effects either. 
These have to be corrected for separately. 

\section{Parametrizing imperfections} \label{sec:ParaImp}

Given the cylindrical symmetry of the ideal Penning trap, it is only natural to emulate the ideal electrostatic potential with electrodes of the same symmetry. 
Holes for injection or ejection, and split electrodes for excitation or signal pickup may violate this ideal, but nevertheless a major component of electrostatic imperfections is expected to be cylindrically-symmetric. 
The same holds true for the distortion of the magnetic field caused by the susceptibility of these trap electrodes, because solenoid magnets intrinsically possess a great deal of cylindrical symmetry, too. 
Moreover, the particle averages over some departures from cylindrical symmetry as it orbits around the center of the trap.  
Given the prevalence of (effective) cylindrical symmetry in the Penning trap, we present an explicit parametrization of such electric and magnetic imperfections in the following two subsections.

\subsection{Electric imperfections} \label{subsec:ParaElecImpec}

In free space, the electrostatic potential~$\Phi$ has to fulfill the Laplace equation~$\Delta \Phi = 0 $, where $\Delta = \mynabla\cdot \mynabla $ is the Laplace operator.
As a Dirichlet problem, the electrostatic potential~$\Phi$ in the Penning trap is then obtained by solving the Laplace equation for the boundary conditions imposed by the trap electrodes. 
Note that by assuming the volume enclosed by the electrodes to be charge-free, we have neglected ion--ion interactions for multiple particles. 
Even for a single trapped ion, its image charge induced on the electrodes acts back on that very same particle. 
This image-charge shift has been treated perturbatively before~\cite{vanDyck1989, porto2001, sturm2013} and is ignored in this paper. 

Cylindrically-symmetric solutions of the Laplace equation which do not diverge at the origin can be written in the form
\begin{gather}
\Phi_{\myinn}(r,\theta) = C_{\myinn}\frac{V_0}{2d^{\myinn}} r^{\myinn}P_{\myinn}\big(\cos(\theta)\big) \label{eq:PhiN-rTheta}
\end{gather}
in spherical coordinates, where $\myinn$ is a non-negative integer and $P_{\myinn}$ a Legendre polynomial. 
The prefactor is in line with our definition of the quadrupole potential ($\myinn=2$) in \myEqua{eq:IPT-quadPot}. 
The full potential~$\Phi$ is then given by a sum over the fundamental solutions~$\Phi_{\myinn}$.
These solutions possess cylindrical symmetry because the azimuth angle~$\phi$, which is the same for spherical and cylindrical coordinates, is absent.  
Despite being ill-suited for the treatment of cylindrically-symmetric imperfections, we have shown the solution in spherical coordinates because it stands as the standard parametrization in the literature. 
Therefore, we wanted to define our coefficients~$C_{\myinn}$ accordingly.%
\footnote{Note that there is some disaccord about whether to normalize coefficients with $\myinn$ odd to the characteristic trap dimension~$d$ or to the distance from the center of the trap to the endcap electrode~\cite{gabrielse1984cpt, gabrielse1984ddt, gabrielse1989}.
Moreover, odd coefficients are usually combined with an asymmetrically applied voltage since the ideal Penning trap has perfect reflection symmetry about the $xy$-plane. 
However, the exact form of odd coefficients is irrelevant here because we will find that the imperfections associated with them do not result in a first-order frequency-shift.}

Nevertheless, cylindrical imperfections are treated best in cylindrical coordinates $\rho = r \sin(\theta)$ and $z = r \cos(\theta)$. 
To this end, we have to determine the coefficient $\myank{\myinn}{k}$ in%
\footnote{%
The solution in cylindrical coordinates must be an even function of $\rho$, because neither $r = \sqrt{\rho^2+z^2}$ nor $\cos(\theta) = z/\sqrt{\rho^2+z^2}$ depend on the sign of $\rho$. 
Conversely, the solution is symmetric with respect to $z$ for $\myinn$ even, and antisymmetric for $\myinn$ odd. }%
\begin{gather}
r^{\myinn}P_{\myinn}\big(\cos(\theta)\big) \equiv \sum_{k=0}^{\left\lfloor\myinn/2\right\rfloor} \myank{{\myinn}}{k}\; z^{\myinn-2k}\,\rho^{2k} 
 \label{eq:rnPn-zRho} \myGSAbstand .
\end{gather}
The upper limit of the sum is given by the floor function%
\footnote{%
This definition of the floor function is only fine for integer arguments, which  suffices for our purpose. 
Generally speaking, taking the floor function of a real number $x$ yields the largest integer not greater than $x$.}%
\begin{gather}
\left\lfloor\frac{\myinn}{2}\right\rfloor = \begin{cases} \frac{\myinn}{2} & \text{if } \myinn \text{ is even,}\\
\frac{\myinn-1}{2} & \text{if } \myinn \text{ is odd.} 
\end{cases}
\label{eq:IntFloorFunction}
\end{gather}
When applying the Laplace operator to \myEqua{eq:rnPn-zRho}, the left-hand side vanishes by design. 
Applying the Laplace operator
\begin{gather}
\Delta = \frac{1}{\rho}\frac{\partial}{\partial \rho}\left(\rho\frac{\partial}{\partial \rho}\right) + \frac{1}{\rho^2}\frac{\partial^2}{\partial \phi^2}+ \frac{\partial^2}{\partial z^2} 
\end{gather}
in cylindrical coordinates to the right-hand side then yields the recursive relationship
\begin{gather}
\myank{\myinn}{k+1} = -\frac{(\myinn-2k)\,(\myinn-2k-1)}{2^2\,(k+1)^2} \, \myank{\myinn}{k} \label{eq:aRecursive} \myGSAbstand.
\end{gather}
It is satisfied by the explicit expression
\begin{gather}
\myank{\myinn}{k} = \frac{(-1)^k}{2^{2k}}\frac{\myinn!}{(\myinn-2k)!\,(k!)^2} 
\label{eq:ank} \myGSAbstand .
\end{gather}
The recursive relationship~\eqref{eq:aRecursive} does not fix the absolute value of $\myank{\myinn}{k}$;
 \myEqua{eq:rnPn-zRho} does because $P_{\myinn}(1) = 1$, and therefore $\myank{\myinn}{0}=1$. 

\subsection{Magnetic imperfections}
For constant electric fields, the Maxwell equation near the center of the trap, where there are no field-coils carrying currents, reads $\mynabla \times \myvec{B}=  0 $.
Since the curl of the gradient of a scalar function vanishes, the magnetic field can be derived from a scalar potential~$\Psi$ via $\myvec{B} = - \mynabla \Psi$. 
Because the  magnetic field naturally has no sources ($\mynabla \cdot \myvec{B}= 0$), the scalar potential has to fulfill the Laplace equation $\Delta \Psi = 0$. 
In analogy to the electrostatic potential~$\Phi$, we write the fundamental solutions which do not diverge at the origin as
\begin{gather}
\Psi_{\myinn}(r,\theta) = -\frac{B_{\myinn}}{\myinn+1}\, r^{\myinn+1} P_{\myinn+1}\big(\cos(\theta)\big) \label{eq:Psi-rnPn}\myGSAbstand .
\end{gather}
Here $B_\myinn$ defines the strength of the contribution. 
Unlike $C_\myinn$, the parameter~$B_\myinn$ is not dimensionless; 
it has the dimension of magnetic field strength times $(\text{length})^{-\myinn}$. 
The total magnetic field is given by a sum over the individual contributions~$\myvec{B}_\myinn$.

By taking the negative gradient
\begin{align}
\myvec{B}_{\myinn}(\rho, z) &= -\frac{\partial\Psi_{\myinn}(\rho,z)}{\partial z} \, \myeuv{z} - \frac{\partial\Psi_{\myinn}(\rho,z)}{\partial \rho} \, \myeuv{\rho}\\
&= B_{\myinn}^{(z)}(\rho, z) \, \myeuv{z}+ B_{\myinn}^{(\rho)}(\rho, z) \, \myeuv{\rho}\myGSAbstand ,
\end{align}
we relate the scalar potential~$\Psi_{n}$ to the axial and radial components of the additional magnetic field. 
Note the slight lapse in notation: 
$\myvec{B}_{\myinn}$ describes a magnetic field, whereas $B_{\myinn}$ is a coefficient that is not equal to the magnitude of that field (unless for $\myinn=0$). 
The axial component 
\begin{align}
B_{\myinn}^{(z)}(\rho, z) &=B_{\myinn}\sum_{k=0}^{\left\lfloor\myinn/2\right\rfloor} \myank{{\myinn}}{k}\; z^{\myinn-2k}\, \rho^{2k} 
\label{eq:BnAxial}
\end{align}
has the same spatial dependence as the electric potential associated with~$C_{\myinn}$. 
The radial component is
\begin{align}
B_{\myinn}^{(\rho)}(\rho, z) &= B_{\myinn} \sum_{k=1}^{\left\lfloor\frac{{\myinn}+1}{2}\right\rfloor}\myatnk{{\myinn}}{k}\; z^{{\myinn}-2k+1}\,\rho^{2k-1} 
\label{eq:Bn-rad}
\end{align}
with the coefficient
\begin{align}
\myatnk{{\myinn}}{k} &= \frac{(-1)^{k}\, k}{ 2^{2k-1}}\frac{{\myinn}!}{({\myinn}-2k+1)!\left(k!\right)^2} 
\label{eq:myantk} \myGSAbstand .
\end{align}
Note that all powers of $\rho$ in $B_{\myinn}^{(\rho)}$ are odd, whereas the powers of $z$ have the opposite parity of $\myinn$. 
Thus, the radial magnetic field is symmetric with respect to $z$ for $\myinn$ odd, and antisymmetric for $\myinn$ even. 

\section{Theoretical framework} \label{sec:TheoFrame}

The real Penning trap is typically a very good approximation of the ideal one. 
Meticulous care is taken manufacturing and assembling the trap electrodes, and in some cases mechanical devices for in-situ alignment of the trap with respect to the magnetic field are installed. 
Moreover, correction electrodes allow for the tuning of lower-order electric imperfections~\cite{vanDyck1976}, while shimming and correction coils achieve the same for magnetic imperfections~\cite{vanDyck1986}. 
Additionally, the stored particle is cooled and by virtue of its small motional amplitudes mainly susceptible to lower-order imperfections. 
Nevertheless, with a relative single-shot  resolution as high as  $10^{-10}$ for $\myomp$~\cite{vanDyck2006, mount2011}, even small frequency-shifts become important. 

The combination of small imperfections with benign consequences is the ideal domain of perturbation theory, which builds the solution bottom-up, order by order, rather than top-down. 
We recall that we have already solved the problem in the absence of imperfections in \mySec{sec:IdealPT}. 
This result will serve as the zeroth-order input. 
Before we get to our formulation of perturbation theory for the specific problem, we present two important trigonometric identities.

\subsection{Powers of cosine}

As one might expect, inserting the solution from \myEquas{eq:x0}--\eqref{eq:z0} for the trajectory in the ideal Penning trap into the higher-order terms brought about by the imperfections will result in powers of trigonometric functions. 
We shall see that we need to analyze the frequency components of such a term. 
By writing $2\cos(\omega t) = \exp(\myi \omega t) + \exp(-\myi \omega t)$, the straightforward application of binomial expansion yields
\begin{align}
\left[\cos(\omega t)\right]^{2n}
&= \frac{1}{2^{2n}}\left[\binom{2n}{n} + 2\sum_{j=1}^{n}\binom{2n}{n-j} \cos(2j\omega t)\right]  \label{eq:cos2N}
\end{align}
with the binomial coefficient defined as
\begin{gather}
\binom{n}{k} = \begin{cases}\dfrac{n!}{k!\,(n-k)!} & \text{if } 0\le k \le n,\\ 0 &\text{otherwise}.\end{cases} \label{eq:BinoDef}
\end{gather}
Although the latter case is prevented by the limits of the sum in \myEqua{eq:cos2N}, we stress that it will become important later on. 
The general formulas for the frequency-shifts given in this paper must be evaluated accordingly. 
By defining the binomial coefficient to vanish for negative $k$ and $n$, as well as for $k>n$, we will be able to write the formulas for the first-order frequency-shifts in a more systematic way. 
In particular, the binomial coefficient has all exceptions covered without using different summation limits for the sums that will show up. 

For odd powers, we obtain
\begin{align}
\left[\cos(\omega t)\right]^{2n+1}&= \frac{1}{2^{2n}}\sum_{j=0}^{n}\binom{2n+1}{n-j} \cos\big[(2j+1)\omega t\big]  \label{eq:cos2Np1} \myGSAbstand.
\end{align}

Even powers of $\cos(\omega t)$ result in a constant term and higher harmonics at even multiples of the fundamental frequency~$\omega$. 
Decomposing the odd powers of $\cos(\omega t)$ results in an oscillatory term at the fundamental frequency~$\omega$ and in higher harmonics at odd multiples thereof. 
However, there is no constant term. 
We shall see that this difference between even and odd powers of an oscillatory term goes a long way in understanding why the imperfections associated with $B_\myinn$ and $C_\myinn$ do not give rise to a first-order frequency-shift for $\myinn$ odd.

Before we proceed, we introduce the piece of notation $\myfComp{\cdots}{\omega}$, which extracts the component at the frequency~$\omega$ from the argument in angle brackets. 
In particular, we have
\begin{gather}
\myfComp{\big[\cos(\omega t)\big]^{2n}}{\myfZero} = \frac{1}{2^{2n}}\binom{2n}{n} = \frac{(2n)!}{2^{2n}(n!)^2} \label{eq:cos2NConst}
\end{gather}
for the constant component and
\begin{align}
\myfComp{\big[\cos(\omega t)\big]^{2n+1}}{\omega} &= \binom{2n+1}{n}\frac{\cos(\omega t)}{2^{2n}} = \frac{(2n+2)!\cos(\omega t)}{2^{2n+1}\,[(n+1)!]^{2}} \label{eq:cos2Np1FundFreq}
\end{align}
for the component at the fundamental frequency~$\omega$. 
Note that, in addition to the amplitude, the oscillatory term is recovered, too. 

\subsection{Perturbation theory} \label{subsec:PertTheory}

While we have solved the classical equations of motion exactly for the ideal Penning trap, general analytic solutions are not available for the ion dynamics in the case of cylindrically-symmetric imperfections. 
As remarked by~\cite{kretzschmar2012icr}, already the low-order anharmonic electric imperfection present enormous difficulties~\cite{horvarth1998, lara2004}. 
Despite the richness of non-linear phenomena, insofar as they do not cause instability, the main concern for Penning-trap mass spectrometry has always been the frequency-shift that goes along, because measuring frequencies is paramount. 
We will briefly touch on motionally-induced dynamics at the end of this subsection. 

With a general solution somewhere between impracticable and impossible, the treatment of anharmonic frequency-shifts in a Penning trap has always been the domain of perturbation theory. 
The calculation of relativistic corrections for an electron in an ideal Penning trap~\cite{graeff1969} ranks among the first applications. 
The first comprehensive summary of frequency-shifts caused by the imperfections of a real Penning trap is given in~\cite{brown1986gtp}, including the effects of the two cylindrically-symmetric electric and magnetic imperfections associated with $C_4$ and $B_2$. 
It is probably because of a long-standing history of measurements on electrons and positrons~\cite{schwinberg1981} on which quantization can be observed that part of the calculation was performed in a quantum-mechanical framework of the anharmonic oscillator. 
Quantum numbers were then related to the energies in the respective modes. 
The result was later picked up on in~\cite{bollen1990} in order to express the shifts as a function of the motional amplitudes, and the same method was applied for $C_6$. 
The quantum-mechanical starting point is overkill for an online-trap working with heavy particles cooled by buffer-gas, and classical methods were used to derive the shift to the radial frequencies in~\cite{gerz1990} for $C_4, C_6$ and $B_2$, albeit named differently. 

Whereas the aforementioned publications remain vague about the actual calculation, showing intermediate steps at best, the prescription for the use of classical Hamiltonian perturbation theory has been outlined and applied in~\cite{kretzschmar1990, kretzschmar1992}. 
Drawing heavily from~\cite{thompson2003, rainville2003}, we opt for a different formalism based on the physical interpretation of the additional forces that result from the higher-order terms.

\paragraph{The one-dimensional anharmonic oscillator}

The train of thought behind the implementation of perturbation theory for the specific problem of calculating first-order frequency-shifts in this paper is illustrated best by exposing the shortcomings of a simple-minded, straightforward approach on the one-dimensional anharmonic oscillator
\begin{gather}
\ddot{z} + \omega_z^2 z = - \varepsilon \myCnfactor{\myinn{}} z^{\myinn-1}  \text{ , where } \myCnfactor{\myinn} = \frac{\myinn\omega_z^2}{2C_2d^{\myinn-2}} \myGSAbstand . \label{eq:1dAnharmOsc}
\end{gather}
The particular choice of the parameter~$\myCnfactor{\myinn}$ is well-motivated. 
By setting the amplitudes of the radial modes to zero, this axial equation of motion results from a Penning trap with an additional contribution~$\Phi_\myinn$ to the electrostatic potential. 
We have made the substitution $\varepsilon = C_\myinn$ for the perturbation parameter.

Since the solution for $\varepsilon = 0$ is known and $\varepsilon$ is typically a small parameter, the ansatz of the solution as a power series 
\begin{gather}
z(t) = z_0(t) + \varepsilon z_1(t)+ \varepsilon^2 z_2(t) + \cdots \label{eq:ZPowerSeriesAnsatz}
\end{gather}
of the perturbation parameter $\varepsilon$ seems to suggest itself. 
The $z_i(t)$ are unknown functions, which need to be determined. 
As usual, the power-series solution is inserted into the original problem of \myEqua{eq:1dAnharmOsc} and subsequently evaluated order by order. 
For the zeroth-order contribution, we have 
\begin{gather}
z_0(t) = \hat{z}_0\cos(\myPhiT{z}) \label{eq:z0t-dump} \myGSAbstand , 
\end{gather}
which is the solution of the one-dimensional harmonic oscillator as was the case for the axial motion in the ideal Penning trap shown in \myEqua{eq:z0}.

By collecting all terms of first-order in $\varepsilon$, we arrive at the differential equation
\begin{gather}
\ddot{z}_1(t) + \omega_{z}^2 z_1(t) =-\myCnfactor{\myinn{}} \big[z_0(t)\big]^{\myinn-1} \label{eq:z1EQM}
\end{gather}
for the first-order contribution~$z_1(t)$. 
Let us now choose $\myinn=2n$ in order to illustrate the problem of the simple-minded power-series ansatz. 
This is a generic problem; 
it would also show up for odd $\myinn$, albeit one order later. 
\myEqua{eq:cos2Np1FundFreq} tells us that from inserting $z_0(t)$ into the right-hand side of \myEqua{eq:z1EQM} we get one oscillatory term at the fundamental frequency and other terms at odd higher harmonics.
Whereas the terms at the higher harmonics lead to motional sidebands in $z_1(t)$, the term 
\begin{gather}
\myfComp{\big[z_0(t)\big]^{2n-1}}{\myomz} = \frac{\hat{z}_0^{2n-1}}{2^{2n-1}}\frac{(2n)!}{(n!)^2} \cos(\myPhiT{z}) \label{eq:oneD-AnharmOscSecTerm}
\end{gather}
right at the fundamental frequency is devastating because it drives the undamped harmonic oscillator for $z_1(t)$ on resonance, eventually leading to a linear growth of the amplitude of $z_1(t)$. 
Even though $z_1(t)$ is suppressed by a factor of $\varepsilon$, the emergence of such a secular term is unphysical because it violates the conservation of energy, and it clearly goes against our expectation of periodicity despite imperfections. 

The pathological effect of secularity is typically removed by introducing multiple (time) scales, a strained variable of time or by developing the frequency as a power-series, too. 
However, we note that all this effort is unnecessary if we are willing to settle for first-order frequency-shifts, provided we understand the message of resonant terms. 

We begin by conceding that our initial ansatz was too restrictive because it did not incorporate frequency as a quantity in its own right and hence lacked the adequate means of describing frequency-shifts in a natural way. 
When the zeroth-order solution~$z_0(t)$ from \myEqua{eq:z0t-dump}, which we believed to be the most important contribution to the overall solution $z(t)$, turned out to oscillate at the unperturbed frequency, we should have questioned the ansatz, given that a frequency-shift was the dominant effect we expected out of imperfections. 
The first order alone cannot set this shortcoming right and diverges. 
If this resonant coupling between different orders of perturbation theory is carried through all orders, the first-order frequency-shift is recovered~\cite{bender1978}.

However, the first-order frequency-shift is related to the resonant term in a much more direct manner based on physical grounds. 
Looking back at our power-series ansatz in \myEqua{eq:ZPowerSeriesAnsatz}, we iterate that there is a hierarchy of contributions to the trajectory. 
We want the zeroth order to be the most important one. 
From \myEqua{eq:z1EQM}, we conclude that the leading contribution to the additional force caused by the imperfection comes from the zeroth-order contribution $z_0(t)$ to the trajectory. 
This is not surprising since the additional force is of first-order in the perturbation parameter~$\varepsilon$ right from the beginning. 
In other words, as the particle goes along $z_0(t)$, it experiences additional forces, most of which are non-resonant and hence only cause small motional sidebands. 
The term from \myEqua{eq:oneD-AnharmOscSecTerm} which looks like a resonant drive-term in \myEqua{eq:z1EQM} actually means that one part of the additional forces has just the same dependence as the force that gave rise to the original motion~$z_0(t)$. 
Therefore, the resonant term does not turn the amplitude into a dynamical quantity; 
it changes the frequency because---effectively---it adds coherently to the main force. 
Consequently, we must allow for a change of the frequency of $z_0(t)$ and  
\begin{gather}
\myZOT{z}(t) = \hat{z}\cos(\myPhiTw{z}) \label{eq:z-ZOT}
\end{gather}
with the perturbed frequency~$\myomzw$ is a better ansatz for the zeroth-order trajectory. 
We have suppressed any indication about the order of $\myZOT{z}$ because we will not have to go beyond zeroth order in the trajectory to calculate the first-order frequency-shift. 

Nevertheless, we stress what this single-minded obsession with frequency-shifts misses out on. 
Sidebands---motional components at higher harmonics of the fundamental frequency---have already been mentioned in conjunction with \myEqua{eq:z1EQM}. 
Note that, for a driven harmonic oscillator, even non-resonant excitation triggers a finite response at the fundamental frequency unless for very special initial conditions. 
If damping is present, only the driven motion survives in the long run, but for the undamped case, the response at the fundamental frequency is here to stay. 
Obviously, the presence of sidebands means that $\hat{z}$ in \myEqua{eq:z-ZOT} is no longer equal to the amplitude of the motion. 
As a more subtle consequence of sidebands, the response at the fundamental frequency also means that $\hat{z}$ is not even equal to the amplitude of the Fourier component at that frequency. 
However, the discrepancy is at least of first order in the perturbation parameter~$\varepsilon$. 
Therefore, we will sloppily continue to refer to $\myAmpZ$ and $\myAmpR{\pm}$ in the case of the perturbed radial modes as the amplitude of that motion in order to avoid the bulky but more exact expression of  ``zeroth-order amplitude of the respective Fourier component of the motion at the perturbed fundamental frequency''.

Our goal is to collect all secularity-inducing terms in the effective equation of motion
\begin{gather}
\ddot{\myZOT{z}}(t) + \omega_z^2(1+\gamma_z) \myZOT{z}(t) = 0 \label{eq:AxialEQM-Eff}\myGSAbstand, 
\end{gather}
from which the perturbed frequency can be read off as
\begin{align}
\myomzw &= \omega_z\sqrt{1+\gamma_z} \approx \omega_z\left(1+\frac{\gamma_z}{2}\right) \label{eq:PaxF} 
\end{align}
with the last approximation assuming $\lvert \gamma_z \rvert \ll 1$. 
By writing the perturbed frequency as $\myomzw = \myomz + \Delta\myomz$, we obtain
\begin{gather}
\frac{\Delta \myomz}{\myomz} = \frac{\gamma_z}{2} \label{eq:DeltaOmz}
\end{gather}
for the first-order axial-frequency shift~$\Delta \myomz$. 

For the one-dimensional anharmonic oscillator presented in \myEqua{eq:1dAnharmOsc}, we would determine the parameter~$\gamma_z$ as
\begin{gather}
\gamma_z \myZOT{z}(t) = \frac{\varepsilon \myCnfactor{\myinn}}{\omega_z^2}\myfComp{  \big[\myZOT{z}(t)\big]^{\myinn-1}}{\myomzw} \myGSAbstand.
\end{gather}
With the help of \myEqua{eq:cos2N}, we note that $\gamma_z=0$ for $\myinn$ odd. 
For $\myinn$ even (given as $\myinn=2n$), the crucial step is to go beyond the result of \myEqua{eq:oneD-AnharmOscSecTerm} by writing the resonant term
\begin{gather}
\myfComp{\big[\myZOT{z}(t)\big]^{2n-1}}{\myomzw} = \frac{\hat{z}^{2n-1}}{2^{2n-1}}\frac{(2n)!}{(n!)^2} \cos(\myPhiTw{z}) = \frac{\hat{z}^{2n-2}}{2^{2n-1}}\frac{(2n)!}{(n!)^2}\, \myZOT{z}(t) \label{eq:myZOT-FF-2nm1}
\end{gather}
as proportional to $\myZOT{z}(t)$. 
Note that the phase of the resonant term is just right to do so.

All factors combined and the substitution $C_{\myinn} = \varepsilon$ undone, the first-order frequency-shift then becomes
\begin{gather}
\frac{\Delta \myomz}{\myomz} = \frac{C_{2n}}{C_2}\frac{n}{2^{2n}}\frac{(2n)!}{(n!)^2}\frac{\hat{z}^{2n-2}}{d^{2n-2}} \label{eq:deltaOmz1d} \myGSAbstand.
\end{gather}
%
\paragraph{Radial modes}

The same strategy as for the axial mode is also applied to the two radial modes. 
We will insert the solutions of the ideal Penning trap 
\begin{align}
\myZOT{x}(t) &= \myZOT{x}_{+}(t) + \myZOT{x}_{-}(t)\myGSAbstand, &\myZOT{y}(t) &= \myZOT{y}_{+}(t) + \myZOT{y}_{-}(t)\myGSAbstand , \label{eq:xyZOT}
\intertext{defining the two abbreviations}
\myZOT{x}_{\pm}(t) &= \myAmpR{\pm}\cos(\myPhiTw{\pm})\myGSAbstand, & 
\myZOT{y}_{\pm}(t) &= -\myAmpR{\pm}\sin(\myPhiTw{\pm}) \label{eq:xypmZOT}
\end{align}
with the same twist: 
we allow for a change of the frequency as indicated by the use of $\myompmw$ instead of the unperturbed frequencies~$\myompm$ in the original solution shown in \myEquas{eq:x0} and~\eqref{eq:y0}. 
We will use the abbreviation~$\myPhiTotw{i}$ as defined in \myEqua{eq:xit} accordingly.

Resonant terms at the frequency~$\myompmw$ are then absorbed in the parameters~$\beta_\pm$ and $\gamma_\pm$ in order to yield the effective radial equations of motion
\begin{gather}
\begin{pmatrix}\ddot{\myZOT{x}}_\pm\\\ddot{\myZOT{y}}_\pm\end{pmatrix} 
=\myomc(1+\beta_\pm) \begin{pmatrix}\dot{\myZOT{y}}_\pm\\-\dot{\myZOT{x}}_\pm\end{pmatrix}+\frac{\myomz^2(1+\gamma_\pm)}{2}\begin{pmatrix}\myZOT{x}_\pm\\\myZOT{y}_\pm\end{pmatrix} 
\label{eq:radEQM-Eff}
\end{gather}
for the zeroth-order trajectory described by $\myZOT{x}_{\pm}(t)$ and $\myZOT{y}_{\pm}(t)$. 
For $\beta_\pm=0$ and $\gamma_\pm = 0$, we recover the radial equations of motions~\eqref{eq:EQM-IPT} of the ideal Penning trap with the free-space cyclotron-frequency~$\myomc$ from \myEqua{eq:fscycf} and the unperturbed axial frequency~$\myomz$ from \myEqua{eq:uptFreqAxial}. 
Note that we have allowed for the fact that the two parameters~$\beta_{\pm}$ and $\gamma_{\pm}$ may be different for the two radial modes. 
We shall see that $\beta_\pm$ is associated with magnetic imperfections, while $\gamma_\pm$ is related to electric imperfections.

By sending $\myomc\to \myomc(1+\beta_\pm)$ and $\myomz^2 \to \myomz^2(1+\gamma_\pm)$ in \myEqua{eq:uptRadFreq}, we use a Taylor expansion around the unperturbed case of $\beta_\pm=0$ and $\gamma_\pm=0$, in order to calculate the first-order frequency-shift as
\begin{align}
\tilde{\omega}_\pm &= \omega_\pm + \frac{\partial \omega_\pm}{\partial \myomc} \myomc\beta_\pm + \frac{\partial \omega_\pm}{\partial \myomz^2} \myomz^2\gamma_\pm + \cdots \label{eq:RadFoFsTaylor}\\
&= \omega_\pm \underbrace{\pm \frac{\omega_\pm \myomc}{\myomp - \myomm}\beta_\pm \mp \frac{\myomp\myomm }{\myomp - \myomm}\gamma_\pm}_{\Delta \myompm} + \cdots \label{eq:RadFoFs} \myGSAbstand.
\end{align}
With the help of \myEqua{eq:omzq-omrad}, we have expressed $\myomz^2$ as a product of the two radial frequencies. 
Note that $\gamma_\pm$ describes a change of the effective $\myomz^2$ in \myEqua{eq:radEQM-Eff}. 
Taking the derivative with respect to $\myomz^2$ instead of $\myomz$ is not a typo. 
Moreover, $\myomz^2(1+\gamma_\pm)$ is not to be confused with an actual axial frequency squared; 
there is only one, and it is given by $\myomz^2(1+\gamma_z)$ in \myEqua{eq:AxialEQM-Eff}. 
Similarly, there is only one free-space cyclotron-frequency $\myomc$, with $\myomc(1+\beta_\pm)$ describing an effective frequency rather than a measurable frequency associated with an actual motion.

The divergence at $\myomp=\myomm$ is an artefact of the first derivative rather than a fundamental flaw of perturbation theory. 
\myEqua{eq:uptRadFreq} does not diverge; 
however, it may yield complex frequencies as a consequence of exceeding the limit of stability in the Penning trap. 
For a measurement on the ion-of-interest, an experiment typically has  $\lvert\myomp\rvert\gg\lvert\myomm\rvert$, and the root in the denominator of \myEqua{eq:RadFoFs} is not an issue. 
Lacking experimental relevance, we are not prepared to deal with the effects of near-degeneracy at the brink of stability.

As we have seen, the first-order frequency-shifts are linear in the perturbation parameter by definition, and the shifts add up linearly as long as the next orders of the Taylor expansion in~\myEquas{eq:PaxF} and~\eqref{eq:RadFoFsTaylor} can be neglected. 
From the perspective of a power series solution, these next-order terms in the Taylor series may be considered at least of second order. 
Because of the linearity to first order in the frequency-shift and the linearity of the equations of motion~\eqref{eq:CNewton-EQM}, as well the superposition principle for electric and magnetic fields, we will treat only one higher-order term at a time, resting assured the effects of several terms can be combined afterwards. 
This holds for multiple electrostatic or magnetic imperfections separately as well as the interplay of both kinds of imperfections. 
Consequently, we devote \mySec{sec:FreqShiftC} to the frequency-shifts caused by former and \mySec{sec:FreqShiftB} to the latter. 
In fact, interplay of imperfections is saying too much. 
It is only in second-order that the effects of different imperfections may conspire to produce a frequency-shift in concert.

\paragraph{Spurious motional resonances}

So far, we have assumed that two kinds of terms arise from inserting the zeroth-order solution into the higher-order imperfections: 
non-resonant and totally coherent ones. 
Since the non-resonant terms do not give rise to a first-order frequency-shift, we have ignored them. 
Non-resonant terms for the one-dimensional anharmonic oscillator involved either a constant term or an oscillatory term at higher harmonics of the fundamental frequency. 
We have assumed the fundamental frequency to be high enough for higher harmonics to be considered non-resonant instead of near-resonant. 
As far as the resonant terms are concerned, our experience with the one-dimensional harmonic oscillator has prompted us to believe that these terms are proportional to a component of the zeroth-order solution, always coming back with the same phase as a component of the original motion. 
We shall call these terms truly resonant. 
However, the three eigenmotions in a Penning trap allow for new phenomena.

What would happen if a term had the right frequency to be resonant, but had incurred a phase-shift along the way?  
We will content ourselves with a handwaving explanation for why there is more than static frequency-shifts to consider in this case. 
Suppose, instead of $\myZOT{z}=\hat{z}\cos(\myPhiTotw{z})$ as for the original motion in the one-dimensional anharmonic oscillator, the resonant force resulting from the anharmonic term came with a phase shift~$\varphi$. 
With the identity
\begin{gather}
\cos(\myPhiTotw{z} + \varphi) = \cos(\myPhiTotw{z})\cos(\varphi) -  \sin(\myPhiTotw{z})\sin(\varphi) 
\end{gather} 
we know how to treat the first term as a frequency-shift. 
In order to understand the effect of the second term, we take a look at the work
\begin{gather}
\myd W = - F \myd z = - F \dot{z} \myd t
\end{gather}
performed by a force $F$ along the way $\myd z$. 
With $F\propto \sin(\myPhiTotw{z})$ and $\dot{\myZOT{z}}\propto \sin(\myPhiTotw{z})$, there is a net change of the particle's energy that does not average over one cycle. 
In other words, the amplitude becomes a dynamical quantity. 
This is in stark contrast to a truly coherent force $F\propto \cos(\myPhiTotw{z})$, whose net influence on the energy of the particle vanishes. 
This vindicates the interpretation of truly resonant terms as an addition to the main force rather than an external drive.

Although this simple model gives a first impression of how instabilities come about as a consequence of imperfections, our self-imposed restriction to zeroth-order trajectories prevents us from seeing the full spectrum of spurious mode-coupling for which motional sidebands offer additional possibilities. 
Moreover, with the frequencies depending on the amplitudes, the resonance conditions change as the amplitudes become dynamical. 
In other words, a resonance-condition may lead to its own demise, receding to near-resonant after having changed the amplitude of an eigenmotion. 
However, since all the eigenmodes are treated as undamped, even near-resonant coupling is still likely to have substantial impact.

A Fourier series expansion of the electrostatic potential performed in~\cite{kretzschmar1990} predicts instabilities for 
\begin{gather}
j_+\myompw + j_-\myommw + j_z\myomzw = 0\text{ , where } \lvert j_+\rvert+\lvert j_- \rvert+\lvert j_z\rvert\le \myinn 
\end{gather}
and the $j_i$ are integers---positive, negative or zero. 
The parameter~$\myinn$ describes the order of the imperfection and is the same as in $C_\myinn$. 
The motional resonances and instabilities have been observed on trapped ions~\cite{huebner1997} and electrons~\cite{paasche2003} as a rapid loss of particles from the trap. 
Given these drastic consequences, frequency-shifts become the lesser problem, and there is little we can do about the rest. 
Our method of absorbing truly resonant terms does not cope with dynamical effects. 
Therefore, we will assume that a stable operating point for the trap far away from instabilities has been chosen. 
Spurious resonances will be ignored for the rest of the paper. 

Before we demonstrate how coherence is born naturally from higher-order imperfections without requiring any particular relationship between the eigenfrequencies, we will derive two very important identities. 
The origin of naturally resonant terms is fully understood without these, but the final result would not look as elegant. 

\subsection{Radial ion displacement}
As we have outlined in the previous subsection, we will insert the zeroth-order solution with the provision for a frequency-change. 
Since we will deal with cylindrically-symmetric imperfections, powers of the radial ion displacement squared play an important role. 
By adding the zeroth-order solutions~$\myZOT{x}$ and $\myZOT{y}$ from \myEqua{eq:xyZOT} in quadrature, we obtain
\begin{gather}
\myZOT{\rho}^2 = \myAmpR{+}^2 + \myAmpR{-}^2+ 2\myAmpR{+}\myAmpR{-}\cos(\myPhiTotw{\mybeat})\myGSAbstand \text{ with } \myGSAbstand\myPhiTotw{\mybeat} = \myPhiTotw{+}-\myPhiTotw{-} \label{eq:rhoqTilde}
\end{gather}
for the zeroth-order radial ion displacement squared. 
Note that $\myZOT{\rho}^2$ is not a single-frequency term; 
it possesses a constant contribution and an oscillatory term at the frequency~$\myombw = \myompw-\myommw$. 
Consequently, the frequency-spectrum of $\myZOT{\rho}^{2n}$ has contributions at $0\myombw, 1\myombw, 2\myombw, 3\myombw, \ldots, n\myombw$, with all higher harmonics up to~$n$, not just even or odd multiples of the fundamental frequency~$\myombw$, regardless of whether $n$ is even or odd.

Fortunately, just two frequency-components in $\myZOT{\rho}^{2n}$ share all the load of preserving and creating naturally  resonant terms: 
the constant component and the term oscillating at the frequency~$\myombw$. 
As a common starting point for calculating these two relevant terms, we have 
\begin{align}
\myZOT{\rho}^{2n}&= \big[\myAmpR{+}^2 + \myAmpR{-}^2+ 2\myAmpR{+}\myAmpR{-}\cos(\myPhiTotw{\mybeat})\big]^{n}\\
&= \sum_{j=0}^{n}\sum_{k=0}^{n-j}\binom{n}{j}\binom{n-j}{k} \myAmpR{+}^{j+2k} \myAmpR{-}^{2(n-k)-j}\,\big[2\cos(\myPhiTotw{\mybeat})\big]^{j} \label{eq:cos2N-ZOT}
\end{align}
after applying binomial expansion twice. 
It is at this point that the pathways begin to differ slightly.

Before we continue with the calculation, we remark that all the powers of $\rho$ we will ever need in this paper are actually integer powers of $\rho^2$. 
Therefore, we do not have to bother about performing a frequency-analysis on~$\rho$, which would be much more complicated because of the square root involved.

\paragraph{Constant term}

\begin{figure}
	\centering
		\includegraphics{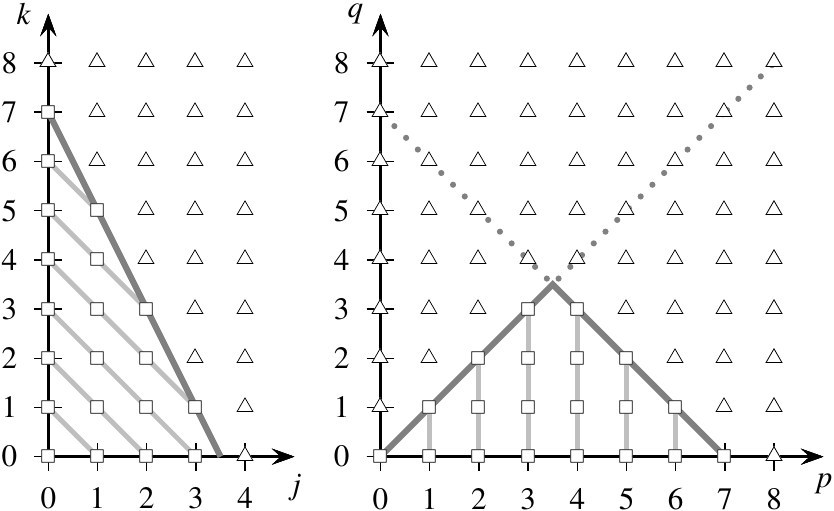}
	\caption{Illustrating the transformation of summation variables in \myEqua{eq:jk-Trafo-pq} for $n=7$. 
	The squares represent combinations of $(j, k)$ and $(p,q)$ that are part of the summation;
	the triangles lie outside the range. 
	In $jk$-space, the dark gray line indicates the upper limit $n-2j$ for $k$. 
	Points of constant $j+k=p$ inside the range of the summation are connected by light gray lines. 
	On such a line, each point is uniquely identified by $j$ aka $q$. 
	As the lines of constant $p$ do not intersect, the coordinates of a point can also be described by $p$ and $j$. 
	For a given $p$, the two constraints $0 \le k \le n-2j$ read $0 \le p-j $ and $p-j \le n-2j $. 
	Thus, we infer that $j \le p$ and $j\le n-p$, in addition to $j\ge 0$. 
	Switching from $j$ to $q$ as the summation variable, the constraints are combined as $0\le q \le \frac{n}{2}-\left\vert \frac{n}{2}-p\right\vert$. 
	The upper limit for $q$ is indicated by the dark gray line in $pq$-space. 
	The dotted lines represent continuations of the upper limits $q= p$ and $q = n-p$ outside the range of the summation.}
	\label{fig:sum-transform}
\end{figure}

Thanks to \myEquas{eq:cos2N} and~\eqref{eq:cos2Np1}, we know that only even powers of $\cos(\myPhiTotw{\mybeat})$ come with a constant term, and we incorporate this by sending~$j\to 2j$ in \myEqua{eq:cos2N-ZOT}. 
The new upper limit in the sum over $j$ is then given by the floor function defined in \myEqua{eq:IntFloorFunction}.
With the result from \myEqua{eq:cos2NConst} for the non-oscillatory component, we have
\begin{align}
\myfComp{\myZOT{\rho}^{2n}}{\myfZero}&= \sum_{j=0}^{\left\lfloor \frac{n}{2}\right\rfloor} \sum_{k=0}^{n-2j}\binom{2j}{j}\binom{n}{2j}\binom{n-2j}{k} \myAmpR{+}^{2(j+k)} \myAmpR{-}^{2(n-j-k)} \myGSAbstand.
\end{align}
Transforming the summation variables as illustrated in \myFig{fig:sum-transform} according to
\begin{gather}
\sum_{j=0}^{\left\lfloor\frac{n}{2}\right\rfloor}\sum_{k=0}^{n-2j} f_n(j, k) = \sum_{p=0}^{n}\sum_{q=0}^{\frac{n}{2}-\left\vert \frac{n}{2}-p\right\vert} f_n(j=q, k= p-q) \label{eq:jk-Trafo-pq}
\end{gather}
with the substitution for $j$ and $k$ as indicated in the argument of a generic function~$f_n(j,k)$ leaves us with
\begin{align}
\myfComp{\myZOT{\rho}^{2n}}{\myfZero} &= \sum_{p=0}^{n}\sum_{q=0}^{\frac{n}{2}-\left\vert \frac{n}{2}-p\right\vert} \binom{2q}{q} \binom{n}{2q} \binom{n-2q}{p-q} \myAmpR{+}^{2p} \myAmpR{-}^{2(n-p)} \myGSAbstand. 
\end{align}
Note that the sum over $q$ entirely determines the coefficient of $\myAmpR{+}^{2p}\myAmpR{-}^{2(n-p)}$ and the sum can be evaluated independently. 
Even more pleasantly, the summation can be executed by hand. 
By rearranging the triple product of binomial coefficients as
\begin{align}
 \binom{2q}{q} \binom{n}{2q} \binom{n-2q}{p-q} 
 &= \binom{n}{p} \binom{p}{q} \binom{n-p}{q} 
\end{align}
and by invoking Vandermonde's identity
\begin{gather}
\sum_{q=0}^{r} \binom{n_1}{q}\binom{n_2}{r-q} = \binom{n_1+n_2}{r} \label{eq:VanderIdentity} \myGSAbstand, 
\end{gather}
the final result
\begin{align}
\myfComp{\myZOT{\rho}^{2n}}{\myfZero} 
&= \sum_{p=0}^{n}\left[\binom{n}{p}\right]^2 \myAmpR{\pm}^{2p} \myAmpR{\mp}^{2(n-p)} \label{eq:rhoqN-DC}
\end{align}
contains one coefficient and only one sum. 
This sum includes all the combinations of $\myAmpR{\pm}^{2p}\myAmpR{\mp}^{2(n-p)}$ and cannot be reduced further because the two amplitudes represent actual degrees of freedom of the experiment. 
We have written $\myAmpR{\pm}$ in order to emphasize that the result is symmetric with respect to the two amplitudes as expected because $\myZOT{\rho}^2$ from \myEqua{eq:rhoqTilde} has the same symmetry. 
We will find it convenient to choose either of the two combinations.

\paragraph{Oscillatory component at $\myombw$}
The starting point for calculating the component at the frequency~$\myombw$ is \myEqua{eq:cos2N-ZOT}, just like for the constant term. 
As evidenced by \myEquas{eq:cos2N} and~\eqref{eq:cos2Np1}, the oscillatory term is present only for odd powers of $\cos(\myPhiTotw{\mybeat})$. 
Therefore, we substitute the summation variable $j\to 2j+1$. 
Using~\myEqua{eq:cos2Np1FundFreq} to determine the term at the frequency~$\myombw$, the intermediate result is
\begin{align}
\myfComp{\myZOT{\rho}^{2n}}{\myombw} &= 2\frac{\myAmpR{+}}{\myAmpR{-}}\cos(\myPhiTotw{\mybeat})\sum_{j=0}^{\left\lfloor \frac{n-1}{2}\right\rfloor}\sum_{k=0}^{n-2j-1} c_{n}(j,k)\, \myAmpR{+}^{2(j+k)}\myAmpR{-}^{2(n-j-k)}
\end{align}
with the coefficient
\begin{align}
c_{n}(j,k) &= \binom{n}{2j+1} \binom{n-2j-1}{k} \binom{2j+1}{j}  \myGSAbstand.
\end{align}
The upper limits of the two sums are not yet right to apply the transformation from \myEqua{eq:jk-Trafo-pq} directly. 
Since $c_n(j, n-2j)=0$ as the second binomial coefficient vanishes according to its definition in \myEqua{eq:BinoDef}, we shift the upper limit of the sum over $k$ up by one to $n-2j$. 
Concerning the sum over $j$, we note that $\lfloor\frac{n-1}{2}\rfloor = \lfloor\frac{n}{2}\rfloor $ for $n $ odd. 
If $n$ is even, the first binomial coefficient vanishes for $2j = n$. 
Therefore, we are allowed to increase the upper limit of the sum over $j$ to $\lfloor\frac{n}{2}\rfloor$ and we can use the same transformation of summation variables as for the constant component before, which results in
\begin{align}
\myfComp{\myZOT{\rho}^{2n}}{\myombw}&=   2\frac{\myAmpR{+}}{\myAmpR{-}}\cos(\myPhiTotw{\mybeat}) \sum_{p=0}^{n}\sum_{q=0}^{\frac{n}{2}-\left\vert \frac{n}{2}-p\right\vert} c_n(q, p-q)\, \myAmpR{+}^{2p}\myAmpR{-}^{2(n-p)} \label{eq:ChapD-rhoqFund1}
\end{align}
with the coefficient
\begin{align}
c_{n}(q,p-q) &= \binom{n}{2q+1} \binom{n-2q-1}{p-q} \binom{2q+1}{q} = \binom{n}{p}\binom{p}{q}\binom{n-p}{q+1} \myGSAbstand.
\end{align}
After rearranging the triple product of binomial coefficients as shown above, Vandermonde's identity~\eqref{eq:VanderIdentity} allows to execute the sum over $q$. 
This final step yields 
\begin{align}
\myfComp{\myZOT{\rho}^{2n}}{\myombw} &=
\frac{2\myAmpR{\pm}}{\myAmpR{\mp}}\cos(\myPhiTotw{\mybeat})\sum_{p=0}^{n}\binom{n}{p}\binom{n}{p+1}\, \myAmpR{\pm}^{2p}\myAmpR{\mp}^{2(n-p)} \label{eq:rhoqN-FF}
\end{align}
for the oscillatory component at the frequency~$\myombw$. 
Note that we could lower the upper limit of the sum over $p$ to $n-1$ as the second binomial coefficient vanishes for $p=n$. 
However, having the same limits as the sum in \myEqua{eq:rhoqN-DC} will help combine variations of the two sums later on. 
Once again, we stress that the result is symmetric with respect to the amplitudes of the two radial modes. 
We underline this by showing both combinations~$\myAmpR{\pm}$, which we will capitalize on to treat the frequency-shifts for both radial modes simultaneously. 

\section{Frequency-shifts caused by electric imperfections} \label{sec:FreqShiftC}
Recalling the parametrization of cylindrically-symmetric imperfections of the electrostatic potential in \mySubsec{subsec:ParaElecImpec}, the additional electric field caused by one higher-order contribution~$\Phi_\myinn$ from \myEqua{eq:PhiN-rTheta} is
\begin{align}
\myvec{E}_{\myinn}(\rho, z) &= -\frac{\partial\Phi_{\myinn}(\rho,z)}{\partial z} \, \myeuv{z} - \frac{\partial\Phi_{\myinn}(\rho,z)}{\partial \rho} \, \myeuv{\rho}\\
&= E_{\myinn}^{(z)}(\rho, z) \, \myeuv{z}+ E_{\myinn}^{(\rho)}(\rho, z) \, \myeuv{\rho}
\end{align}
with the axial component 
\begin{align}
E_{\myinn}^{(z)} &= 
  -C_{\myinn}\frac{V_0}{2d^{\myinn}}\sum_{k=0}^{\left\lfloor\myinn/2\right\rfloor} \myank{\myinn}{k}\  (\myinn-2k)\,z^{\myinn -2k - 1} \rho^{2k} \label{eq:EnAxial}
\end{align}
and the radial component 
\begin{align}
E_{\myinn}^{(\rho)} &= 
 -C_{\myinn}\frac{V_0}{d^{\myinn}}\sum_{k=1}^{\left\lfloor \myinn/2 \right\rfloor} \myank{\myinn}{k}\  k\, z^{\myinn-2k} \rho^{2k-1} \myGSAbstand.
\end{align}
The coefficient~$\myank{\myinn}{k}$  is as defined in \myEqua{eq:ank}.
By expressing the radial unit vector~$\myeuv{\rho}$ in Cartesian coordinates, the two Cartesian  components of the additional radial electric field are found as
\begin{gather}
\begin{pmatrix}E_{\myinn}^{(x)} \\ E_{\myinn}^{(y)} \end{pmatrix} =  \frac{E_{\myinn}^{(\rho)} }{\rho}\begin{pmatrix}x \\ y \end{pmatrix}= -C_{\myinn}\frac{V_0}{d^{\myinn}}\sum_{k=1}^{\left\lfloor\myinn/2\right\rfloor} \myank{\myinn}{k}\  k\;z^{\myinn-2k}\, \rho^{2k-2} \begin{pmatrix}x \\ y \end{pmatrix} \label{eq:Enxy}\myGSAbstand .
\end{gather}

Since the classical Newtonian equation of motion~\eqref{eq:CNewton-EQM} for a charged particle in electric and magnetic fields is linear in both fields, we can simply add the additional electric field $\myvec{E}_{\myinn}$ to the equations of motion~\eqref{eq:EQM-IPT} in the ideal Penning trap in order to arrive at 
\begin{align}
\begin{pmatrix} \ddot{x}\\\ddot{y}\\\ddot{z}\end{pmatrix} 
&=\myomc \begin{pmatrix}\dot{y}\\-\dot{x}\\0\end{pmatrix} + \frac{qV_0C_2}{2md^2}\begin{pmatrix}x\\y\\-2z\end{pmatrix}  + \frac{q}{m}\begin{pmatrix}E_{\myinn}^{(x)}\\E_{\myinn}^{(y)}\\E_{\myinn}^{(z)}\end{pmatrix} \label{eq:RPT-EQM-C}\myGSAbstand.
\end{align}
We will search for frequency-changing terms in the axial and the radial modes separately.

\subsection{Axial mode}
Plugging in the additional axial electric field from \myEqua{eq:EnAxial} into \myEqua{eq:RPT-EQM-C}, the axial equation of motion reads
\begin{gather}
\ddot{z} + \myomz^2 z + \frac{C_{\myinn}}{C_2}\frac{\myomz^2}{2d^{\myinn-2}}\sum_{k=0}^{\left\lfloor\myinn/2\right\rfloor} \myank{\myinn}{k}\, (\myinn-2k)\, z^{\myinn-2k-1} \rho^{2k} = 0 
\end{gather}
with the unperturbed axial frequency~$\myomz$ introduced in \myEqua{eq:uptFreqAxial}.

We insert the zeroth-order solution $\myZOT{z}(t)$ and $[\myZOT{\rho}(t)]^2$ from \myEquas{eq:z-ZOT} and~\eqref{eq:rhoqTilde}, respectively. 
In order to save some space, we will suppress the time-dependence. 
In contrast to  the amplitude-like constant quantities~$\myAmpR{\pm}$ and $\myAmpZ$, the zeroth-order solutions expressed with a tilde are explicitly time-dependent. 

Next, we need to understand how oscillatory terms at the axial frequency~$\myomzw$ are produced. 
Spurious motional resonances aside, the natural mechanism is written as
\begin{gather}
\myfComp{\myZOT{z}^{\myinn-2k-1}\myZOT{\rho}^{2k}}{\myomzw}  = \myfComp{\myZOT{z}^{\myinn-2k-1}}{\myomzw} \myfComp{\myZOT{\rho}^{2k}}{\myfZero} 
\end{gather}
in our notation. 
In words, an oscillatory component at the perturbed axial frequency~$\myomzw$ results from that very component in $\myZOT{z}^{\myinn-2k-1}$, and the component is preserved by mixing it only with the constant contribution in $\myZOT{\rho}^{2k}$, because the oscillatory components in $\myZOT{\rho}^{2k}$ are at multiples of $\myombw= \myompw-\myommw$. 
Note that according to \myEquas{eq:cos2N} and \eqref{eq:cos2Np1} the term $\myZOT{z}^{\myinn-2k-1}$ has a contribution at the perturbed axial frequency~$\myomzw$ only if the exponent~$\myinn-2k-1$ is odd. 
Therefore, $\myinn$ must be even, which we incorporate by writing $\myinn = 2n$. 
The terms for our effective equation of motion~\eqref{eq:AxialEQM-Eff} are then expressed as
\begin{align}
\gamma_z \myZOT{z} 
&= \frac{C_{2n}}{C_2}\frac{1}{d^{2n-2}} \sum_{k=0}^{n-1} \myank{2n}{k}\; (n-k)\myfComp{\myZOT{z}^{2n-2k-1}}{\myomzw} \myfComp{\myZOT{\rho}^{2k}}{\myfZero} \label{eq:C2n-gammaz-z} \myGSAbstand.
\end{align}

Even though we have figured out that coefficients with $\myinn$ odd are not a natural  source of first-order frequency-shifts, we highlight one peculiarity. 
$C_1$ is special among the odd coefficients because its contribution is constant, which translates the center of the axial oscillation. 
In the ideal Penning trap with its homogeneous and therefore translationally invariant magnetic field, $C_1$ does not cause a frequency-shift because $\Phi_1\propto z$ without any dependence on $\rho$ neither changes the curvature of the potential in the axial direction nor affects the electrostatic potential for radial modes at all. 
Moreover, there is no particular location in a homogeneous magnetic field. 

The effect of $C_1$ could be encompassed non-perturbatively in the translated zeroth-order solution
\begin{gather}
\myZOT{z}'(t) = \myZOT{z}(t) - \frac{C_1}{2C_2}d \myGSAbstand.
\end{gather}
However, calculating powers of $\myZOT{z}'(t)$ would be more tedious. 
Nevertheless, the additional terms would at least be of first order in $C_1$, and it is still true that $C_1$ does not give rise to a frequency-shift to first-order. 
Note that since $\Phi_1\propto z$, the radial modes would only be affected via the $C_1$ in $\myZOT{z}'(t)$, which means that the effect is at least of second order because it has to be mediated by a higher-order term, $z$ being absent from the radial equations of motion in the ideal Penning trap. 
Sticking with the original $\myZOT{z}(t)$ still allows for a perturbative treatment of $C_1$, but as we have seen one has to go beyond the friendly confines of first order. 

Alternatively, the electrostatic potential and the magnetic field may be parametrized around the new center of the axial oscillation, where there is no effective~$C_1$. 
Compromising the reflection symmetry about the $xy$-plane results in the proliferation of perturbation parameters with $\myinn$ odd. 
Although these terms do not go along with a first-order frequency-shift and their effects may be considered subordinate, a quantitative treatment requires second order in perturbation theory as before. 

The effect of odd coefficients is more subtle than one might expect at first glance, in large parts because there is no mechanism to ensure that the $C_{2n+1}$ are small with respect to $C_2$. 
For hyperboloidal traps, the shape of the electrodes guarantees that $C_2$ dominates over $C_{2n}$, but the $C_{2n+1}$ cannot be dismissed along these lines. 
When reflection symmetry of the potential is deliberately broken, for instance by introducing an offset voltage on one electrode, the second-order effects of odd coefficients may not necessarily be smaller than the first-order effects of even coefficients. 
Moreover, since typical traps barely resemble plate capacitors,  $C_1$ and $C_3$ are of the same order or magnitude~\cite{gabrielse1984cpt, gabrielse1984ddt, gabrielse1989}, and the two conspire to produce an effective $C_2$, thereby causing a frequency-shift independent of the particle's amplitudes. 
In this case, seemingly anharmonic terms combine to yield a harmonic shift. 
Whereas a change of the effective~$C_2$ affects the individual frequencies, \myEquas{eq:SB-Cyc} and~\eqref{eq:invt} remain valid. 
As the $C_1C_3$-shift is of second order in the perturbation parameters, we will stay true to the title of this paper by ignoring it. 

Nevertheless, we have to warn of mistaking orders for actual relevance. 
Speaking of orders in perturbation theory is simply adding the exponents of all the perturbation parameters in a particular term. 
When different perturbation parameters are compared, higher orders do not necessarily imply lesser importance, although we hope to operate in a regime of imperfections and amplitudes in which the hierarchy is at least maintained for the terms that involve the same set of perturbation parameters. 
In other words, the first-order effect of $B_{2n}$ or $C_{2n}$ is supposed to dominate over its second-order effect. 
For anharmonic shifts, the assumption is verified experimentally by measuring a frequency-shift as a function of the amplitude of an eigenmotion. 
If the second-order effects of $B_{2n}$ or $C_{2n}$ were important, the scaling with amplitude would differ from the first-order prediction. 
However, the second-order effects of odd coefficients may be much harder to disentangle from the first-order effects of even coefficients. 

Having underlined the importance of reflection symmetry in a Penning trap, we return to the truly resonant terms given by \myEqua{eq:C2n-gammaz-z}. 
Using \myEquas{eq:myZOT-FF-2nm1} and~\eqref{eq:rhoqN-DC} to evaluate the two terms in angle brackets, and the definition of $\myank{2n}{k}$ from \myEqua{eq:ank}, the parameter $\gamma_z$ is written as
\begin{align}
\begin{split}
\gamma_z = \frac{C_{2n}}{C_2}\frac{(2n)!}{2^{2n-1}d^{2n-2}}&\sum_{k=0}^{n-1} \frac{(-1)^{k}\, (n-k)}{[k!(n-k)!]^2}\\
 &\;\cdot\sum_{p=0}^{k}\left[\binom{k}{p}\right]^2 \myAmpR{+}^{2p} \myAmpR{-}^{2(k-p)} \myAmpZ^{2(n-k-1)} 
\myGSAbstand.
\end{split}
\end{align}
Expressing the binomial coefficient with factorials according to \myEqua{eq:BinoDef}, the parameter~$\gamma_z$ is related to the first-order frequency-shift
\begin{align}
\frac{\Delta \myomz}{\myomz} &= \frac{C_{2n}}{C_2}\frac{(2n)!}{2^{2n}}\sum_{k=0}^{n-1}\sum_{p=0}^{k} \frac{(-1)^{k}\, (n-k) \, \myAmpR{+}^{2p} \myAmpR{-}^{2(k-p)}\myAmpZ^{2(n-k-1)}}{[(n-k)!\, p! \,(k-p)!]^2 \, d^{2n-2} } \label{eq:Cn-FS-Axial}
\end{align}
with the help of \myEqua{eq:DeltaOmz}. 
For zero amplitude of the radial modes, the only contribution comes from $p=0$ and $k=0$, and we recover the result from \myEqua{eq:deltaOmz1d} for the one-dimensional case as a little crosscheck. 
The more comprehensive benchmark is Equation~(3.57) in~\cite{cptI2005} and it agrees.

Note that the result is symmetric with respect to the amplitudes of the two radial modes. 
If they were swapped, the frequency-shift caused by the electrostatic imperfection would not change. 
Essentially, this symmetry is due to the fact that the coupling of the radial modes to the axial motion is mediated non-resonantly through the amplitudes, but not velocities. 
We shall see that this changes for magnetic imperfections. 
More precisely, via electrostatic imperfections, the axial mode is sensitive to the average of powers of the radial ion displacement.

\subsection{Radial modes} \label{subsec:CRadModes}
By inserting the radial components  of the additional electric field shown in~\myEqua{eq:Enxy} into  \myEqua{eq:RPT-EQM-C}, the radial equations of motion become
\begin{align}
\begin{pmatrix}\ddot{x}\\\ddot{y}\end{pmatrix}   
&=\myomc \begin{pmatrix}\dot{y}\\-\dot{x}\end{pmatrix} + \frac{\myomz^2}{2}
\begin{pmatrix}x \\ y \end{pmatrix}
\left[1  - \frac{C_{\myinn}}{C_2}\frac{2}{d^{\myinn-2}} \sum_{k=1}^{\left\lfloor\myinn/2\right\rfloor} \myank{\myinn}{k}\; k\, z^{\myinn-2k} \rho^{2k-2}\right] 
 \label{eq:CnRadEQM} 
\end{align}
with the unperturbed axial frequency~$\myomz$ from \myEqua{eq:uptFreqAxial}.

Like for the axial mode, we plug in the zeroth-order solutions for the trajectory searching for resonant terms. 
In addition to $\myZOT{z}(t)$ from \myEqua{eq:z-ZOT} and $[\myZOT{\rho}(t)]^2$ from \myEqua{eq:rhoqTilde}, we will need the zeroth-order solution for the two radial coordinates given in \myEqua{eq:xyZOT}. 

It is immediately obvious that $\myZOT{x}_\pm(t)$ in $\myZOT{x}(t)$, and $\myZOT{y}_\pm(t)$ in $\myZOT{y}(t)$ as defined in \myEqua{eq:xypmZOT} stay resonant at $\myompmw$ as long as they are  multiplied only with constant components. 
However, we must resist our temptation to average over the sum in square brackets in \myEqua{eq:CnRadEQM} because the two frequency-components in $\myZOT{x}(t)$ and $\myZOT{y}(t)$ allow for a second mechanism: 
producing resonant terms via mixing. 
From \myEqua{eq:rhoqTilde} we recall that the radial ion displacement squared and powers thereof contain an oscillatory term at the difference frequency~$\myombw= \myompw-\myommw$ of the two radial modes. 
With the standard laws of multiplying trigonometric functions, we have
\begin{align}
\myfComp{\vphantom{x^2}\cos(\myPhiTotw{+}-\myPhiTotw{-})\cos(\myPhiTotw{\pm})}{\myommpw}&= \frac{1}{2}\cos(\myPhiTotw{\mp}) \label{eq:cosBeatcospm} \myGSAbstand ,\\
\myfComp{\vphantom{x^2}\cos(\myPhiTotw{+}-\myPhiTotw{-})\sin(\myPhiTotw{\pm})}{\myommpw}&= \frac{1}{2}\sin(\myPhiTotw{\mp})\label{eq:cosBeatsinpm}\myGSAbstand .
\end{align}
In other words, mixing an oscillatory term at the frequency~$\myombw$ with an oscillatory term at the radial eigenfrequency~$\myompmw$ results in a resonant term at the other radial frequency~$\myommpw$. 
Of course, there is a second term at the frequency $\lvert 2\myompmw-\myommpw \rvert$, which we have ignored because it is generally non-resonant. 
Note that \myEquas{eq:cosBeatcospm} and~\eqref{eq:cosBeatsinpm} mean that the mechanism for producing coherence via mixing is the same for $\myZOT{x}(t)$ and $\myZOT{y}(t)$. 
From now on, we will extract naturally resonant terms from one of the two radial coordinates, knowing that these terms can be described by one single parameter for both coordinates. 
This comes as no surprise since cylindrical symmetry must not favor one radial direction over the other.

In our notation, producing resonant terms in \myEqua{eq:CnRadEQM} is written as
\begin{gather}
\myfComp{\myZOT{z}^{\myinn-2k} \myZOT{\rho}^{2k-2} \myZOT{x}}{\myompmw} = \myfComp{\myZOT{z}^{\myinn-2k}}{0} \myfComp{\myZOT{\rho}^{2k-2} \myZOT{x}}{\myompmw} \myGSAbstand.
\end{gather}
Barring spurious resonances, oscillatory terms at the radial frequency~$\myompmw$ result from the radial motions themselves, and these terms need to be preserved by multiplying them with the time-independent contribution of the axial oscillation. 
According to \myEquas{eq:cos2N} and~\eqref{eq:cos2Np1}, the exponent $\myinn-2k$ has to be even for $\myZOT{z}^{\myinn-2k}$ to have a constant term. 
Therefore, $\myinn$ must be even. 
As for the axial mode before, we incorporate this property by writing~$\myinn=2n$. 

Resonant terms in the radial equations of motion~\eqref{eq:CnRadEQM} then take the form
\begin{align}
\gamma_\pm \myZOT{x}_\pm 
&=-\frac{C_{2n}}{C_2}\frac{2}{d^{2n-2}}\sum_{k=1}^{n} \myank{2n}{k}\; k \myfComp{\myZOT{z}^{2(n-k)}}{\myfZero} \myfComp{\myZOT{\rho}^{2k-2} \myZOT{x}}{\myompmw} 
\end{align}
with the parameter~$\gamma_\pm$ defined in the effective equations of motion~\eqref{eq:radEQM-Eff}.
The constant contribution by the axial motion is derived from \myEqua{eq:cos2NConst}. 
The coefficient~$\myank{2n}{k}$ is defined in \myEqua{eq:ank}.
Oscillatory terms at either radial frequency~$\myompmw$ are produced by the two mechanisms discussed before: 
on the one hand, preserving the resonant term present in $\myZOT{x}_\pm(t)$ by multiplying it with constant terms only; 
on the other hand, mixing the term $\myZOT{x}_\mp(t)$ with the oscillatory component at $\myombw = \myompw-\myommw$ to produce a term proportional to $\myZOT{x}_\pm(t)$. 
In our notation, these mechanisms read
\begin{gather}
\myfComp{\myZOT{\rho}^{2k-2} \myZOT{x}}{\myompmw}
= \myfComp{\myZOT{\rho}^{2k-2}}{\myfZero} \myZOT{x}_{\pm} + \myfComp{\myfComp{\myZOT{\rho}^{2k-2}}{\myombw} \myZOT{x}_{\mp}}{\myompmw} \myGSAbstand.
\end{gather}
Note that there is no double counting because $\myZOT{x}_\pm(t)$ is multiplied with two distinct components of powers of the radial displacement squared. 
Combining $\myZOT{x}_{+}(t)$ with the constant component of $\myZOT{\rho}^{2k-2}$ results in a resonant term at the reduced cyclotron-frequency~$\myompw$; 
mixing $\myZOT{x}_{+}(t)$ with the oscillatory component  of $\myZOT{\rho}^{2k-2}$ at $\myombw$ yields a resonant term at the magnetron frequency~$\myommw$. 
The same holds true for $\myZOT{y}_+(t)$. 
Similarly, the same mechanism is at work for the effect of $\myZOT{x}_-(t)$ and $\myZOT{y}_-(t)$ on $\myommw$ and $\myompw$.

With the help of \myEqua{eq:rhoqN-DC}, we write the first of the two summands as
\begin{align}
\myfComp{\myZOT{\rho}^{2k-2}}{\myfZero} \myZOT{x}_{\pm} &=  \myZOT{x}_\pm \sum_{p=0}^{k-1} \left[\binom{k-1}{p}\right]^2 \myAmpR{\pm}^{2p} \myAmpR{\mp}^{2(k-1-p)} \myGSAbstand.
\end{align}
Using~\myEquas{eq:rhoqN-FF} and \eqref{eq:cosBeatcospm}, the second summand becomes 
\begin{align}
\myfComp{ \myfComp{\myZOT{\rho}^{2k-2}}{\myombw} \myZOT{x}_{\mp}}{\myompmw} 
&= \myZOT{x}_\pm \sum_{p=0}^{k-1}\binom{k-1}{p}\binom{k-1}{p+1}\, \myAmpR{\pm}^{2p}\myAmpR{\mp}^{2(k-1-p)} \myGSAbstand.
\end{align}
We have chosen the combination~$\myAmpR{\pm}$ in \myEqua{eq:rhoqN-FF} such that the factor $\myAmpR{\pm}/\myAmpR{\mp}$ compensates for the factor $\myAmpR{\mp}/\myAmpR{\pm}$ that results from producing $\myZOT{x}_\pm$ from $\myZOT{x}_\mp$ through mixing. 
Recall that $\myZOT{x}_\mp$ comes with $\myAmpR{\mp}$, whereas an additional factor of $\myAmpR{\pm}$ is necessary to create $\myZOT{x}_\pm$ after mixing with the oscillatory component at $\myombw$. 
Accidentally, this choice also means that the exponents of the radial amplitudes~$\myAmpR{\pm}$ are the same in both sums, which allows us to proceed without any transformation of summation variables. 

With the identity
\begin{align}
\binom{k-1}{p} +\binom{k-1}{p+1} &= \binom{k}{p+1} \label{eq:BinoAddId1}
\end{align}
the two results can be combined and---the contribution from the axial motion considered---the coefficient of the resonant terms at $\myompmw$ takes the form
\begin{gather}
\begin{split}
\gamma_\pm  &= -\frac{C_{2n}}{C_2}\frac{(2n)!}{2^{2n-1}}\frac{1}{d^{2n-2}} \sum_{k=1}^{n} \frac{(-1)^{k}k}{[k!(n-k)!]^2 }\\
 &\quad  \cdot \sum_{p=0}^{k-1}\binom{k-1}{p}\binom{k}{p+1} \myAmpR{\pm}^{2p}\myAmpR{\mp}^{2(k-1-p)} \myAmpZ^{2(n-k)} \myGSAbstand.
 \end{split}
\end{gather}
Note that both binomial coefficients are defined by their explicit expression with factorials shown in~\myEqua{eq:BinoDef} within the limits of the sum, and we will consequently replace the binomial coefficients with factorials. 
\myEqua{eq:RadFoFs} relates $\gamma_\pm$ to the first-order frequency-shift of the radial modes and we obtain
\begin{align}
\begin{split}
\Delta \myompm  &= \frac{\pm\myomp\myomm }{\myomp - \myomm} \frac{C_{2n}}{C_2}\frac{(2n)!}{2^{2n-1}}\frac{1}{d^{2n-2}}\\&\quad \cdot  \sum_{k=1}^{n}\sum_{p=0}^{k-1}\frac{(-1)^{k}\,(p+1)\, \myAmpR{\pm}^{2p}\myAmpR{\mp}^{2(k-1-p)} \myAmpZ^{2(n-k)}}{[(n-k)!\,(k-p-1)!\,(p+1)!]^2}   \myGSAbstand.
\end{split} \label{eq:Cn-FS-Radial}
\end{align}
Equation~(3.56) in~\cite{cptI2005} provides a crosscheck and there is agreement.

By transforming the summation variable according to $p\to k-p-1$, the exponents of the two radial amplitudes are swapped and we obtain
\begin{gather} 
\begin{split} \Delta \myompm &= \frac{\pm\myomp\myomm }{\myomp - \myomm} \frac{C_{2n}}{C_2}\frac{(2n)!}{2^{2n-1}}\frac{1}{d^{2n-2}}\\&\quad \cdot  \sum_{k=1}^{n}\sum_{p=0}^{k-1}\frac{(-1)^{k}\,(k-p)\, \myAmpR{\mp}^{2p}\myAmpR{\pm}^{2(k-1-p)} \myAmpZ^{2(n-k)}}{[(n-k)!\,p!\,(k-p)!]^2} \label{eq:Cn-FS-Radial2} \myGSAbstand.
 \end{split} 
\end{gather}
It is now clear to see that the result is not symmetric with respect to the two amplitudes of the radial motions. 
Technically, this imbalance results from producing resonant terms at $\myompmw$ by mixing with $\myZOT{x}_\mp$, which brings in the amplitude~$\myAmpR{\mp}$ of the other radial motion in an asymmetric way. 

\paragraph{The cyclotron sideband}

The shift to the sideband cyclotron-frequency defined in \myEqua{eq:SB-Cyc} is simply calculated by adding the two shifts of the radial modes. 
From \myEquas{eq:Cn-FS-Radial} and~\eqref{eq:Cn-FS-Radial2} we have
\begin{align}
\begin{split} 
\Delta \myomc  &= \frac{\myomp\myomm }{\myomp - \myomm} \frac{C_{2n}}{C_2}\frac{(2n)!}{2^{2n-1}} \frac{1}{d^{2n-2}}\sum_{k=1}^{n} \frac{(-1)^{k}\myAmpZ^{2(n-k)}}{[(n-k)!]^2 }\\
  &\quad \cdot \sum_{p=0}^{k-1}\frac{\myAmpR{+}^{2p} \myAmpR{-}^{2(k-1-p)}}{[p!\,(k-p-1)!]^2}  \left[\frac{1}{p+1}-\frac{1}{k-p}\right]  \myGSAbstand .
  \end{split} \label{eq:Cn-FS-SB}
\end{align}
By substituting the summation variable as $p\to k-p-1$ as before, the exponents of the two radial amplitudes are swapped and we obtain
\begin{align}
\begin{split}
\Delta \myomc   &= -\frac{\myomp\myomm }{\myomp - \myomm} \frac{C_{2n}}{C_2}\frac{(2n)!}{2^{2n-1}} \frac{1}{d^{2n-2}}\sum_{k=1}^{n} \frac{(-1)^{k}\myAmpZ^{2(n-k)}}{[(n-k)!]^2 }\\
  &\quad \cdot \sum_{p=0}^{k-1}\frac{\myAmpR{-}^{2p} \myAmpR{+}^{2(k-1-p)}}{[p!\,(k-p-1)!]^2}  \left[\frac{1}{p+1}-\frac{1}{k-p}\right] \myGSAbstand .
  \end{split} \label{eq:Cn-FS-SB2}
\end{align}
Clearly, the result is antisymmetric with respect to the amplitudes of the radial modes. 
If the two radial amplitudes are swapped, the frequency-shift~$\Delta \myomc$ changes sign. In particular, the shift must vanish for equal radial amplitudes. 
This finding is in line with~\cite{cptI2005, brodeur2012}, where the common root $(\myAmpR{+}^2-\myAmpR{-}^2)$ is factored out in the specific expressions for the first few $C_{2n}$. 
We now understand this as a general feature of cylindrically-symmetric electrostatic imperfections. 

Our interest in swapping the radial amplitudes stems from an experimental motivation. 
Probing the degree of conversion of an initial~$\myAmpR{-}$ into $\myAmpR{+}$ is the standard method for measuring the sideband cyclotron-frequency at online traps~\cite{koenig1995}. 
The amplitude and the duration of the coupling pulse are typically chosen such that there is a full conversion when the frequency of the pulse coincides with the sideband cyclotron-frequency.

Recently, individual frequency-measurements on the two radial modes have been combined~\cite{eliseev2013} to calculate the sideband cyclotron-frequency via \myEqua{eq:SB-Cyc}. 
Essentially, the phase-sensitive measurement of~$\myompmw$ is performed after having excited the amplitude of the corresponding radial mode to $\myAmpR{\pm} = \myAmpR{\myhot}$, whereas the amplitude of the other radial mode is $\myAmpR{\mp} = \myAmpR{\mycool}$ during the evolution time of the phase. 
Assuming equal axial amplitudes for both measurements, the shifts from \myEqua{eq:Cn-FS-Radial} are equal in magnitude but opposite in sign: 
\begin{gather}
\Delta \myomp(\myAmpR{\myhot}, \myAmpR{\mycool}, \myAmpZ) = -\Delta \myomm(\myAmpR{\mycool}, \myAmpR{\myhot}, \myAmpZ) 
\label{eq:PICRcancel} \myGSAbstand.
\end{gather}
The order in both arguments is $(\myAmpR{+}, \myAmpR{-}, \myAmpZ)$. 
Thus, these two shifts cancel in the sideband identity~\eqref{eq:SB-Cyc} for this particular measurement scheme. 
\myEqua{eq:PICRcancel} also confirms that the shift~$\Delta \myomc$ changes sign when the amplitudes of the radial modes are swapped. 

\section{Frequency-shifts caused by magnetic imperfections} 
\label{sec:FreqShiftB}

With the additional magnetic field, the classical Newtonian equations of motion~\eqref{eq:CNewton-EQM}, shown in \myEqua{eq:EQM-IPT} for the ideal Penning trap, become
\begin{align}
\begin{pmatrix}\ddot{x}\\\ddot{y}\\\ddot{z}\end{pmatrix}
&=\myomc \begin{pmatrix}\dot{y}\\-\dot{x}\\0\end{pmatrix} +\frac{\myomc}{B_0}\begin{pmatrix}\dot{y}B_{\myinn}^{(z)}- \dot{z}B_{\myinn}^{(y)}\\ \dot{z}B_{\myinn}^{(x)}- \dot{x}B_{\myinn}^{(z)}\\ \dot{x}B_{\myinn}^{(y)}- \dot{y}B_{\myinn}^{(x)}\end{pmatrix}+ \frac{\myomz^2}{2}\begin{pmatrix}x\\y\\-2z\end{pmatrix} \label{eq:RealBEqMotions} \myGSAbstand,
\end{align}
where the radial magnetic field from \myEqua{eq:Bn-rad} is translated into Cartesian coordinates as
\begin{gather}
\begin{pmatrix}
B_{\myinn}^{(x)} \\B_{\myinn}^{(y)} 
\end{pmatrix}
= \frac{B_{\myinn}^{(\rho)}(\rho, z)}{\rho} 
\begin{pmatrix}
x\\ y
\end{pmatrix} 
= B_{\myinn} \sum_{k=1}^{\left\lfloor\frac{\myinn+1}{2}\right\rfloor}\myatnk{{\myinn}}{k}\, z^{{\myinn}-2k+1}\rho^{2k-2}
\begin{pmatrix}
x\\ y 
\end{pmatrix} \label{eq:Bn-radCart} 
\end{gather}
just like the radial electric field in \myEqua{eq:Enxy}. 

Like for electrostatic imperfections, we will examine the axial mode and the radial modes separately. 
Because the magnetic field couples to velocities, the additional factor of $\myomw_i$ that results from taking the time-derivative of the zeroth-order trajectory makes the treatment more cumbersome, but the mechanism for producing frequency-shifting terms stays the same as for electrostatic imperfections. 
As we will not encounter new phenomena, we will remain brief about the basics.

\subsection{Axial mode} \label{subsec:MagAxialMode}

By inserting the radial magnetic field from \myEqua{eq:Bn-radCart} into the third component of \myEqua{eq:RealBEqMotions}, the axial equation of motion  becomes
\begin{gather}
\ddot{z} + \myomz^2 z - \myomc \frac{B_{\myinn}}{B_0}
\left(\sum_{k=1}^{\left\lfloor\frac{\myinn+1}{2}\right\rfloor}\myatnk{\myinn}{k}\, z^{{\myinn}-2k+1}\rho^{2k-2}\right)\left(\dot{x}y- \dot{y}x\right) = 0 \label{eq:BaxialEQM}\myGSAbstand. 
\end{gather}

As usual, we will plug in the zeroth-order solutions from \myEquas{eq:z-ZOT} and~\eqref{eq:xyZOT}, while still allowing for a frequency-shift. 
We start out by defining and evaluating the abbreviation
\begin{align}
\myZOT{\xi}(t)&= \dot{\myZOT{x}}(t)\myZOT{y}(t)- \dot{\myZOT{y}}(t)\myZOT{x}(t) \\
 &= \myAmpR{+}^2\myompw + \myAmpR{-}^2\myommw + \myAmpR{+} \myAmpR{-}(\myompw+\myommw)\cos(\myPhiTotw{+}-\myPhiTotw{-}) \label{eq:xiZOT}
\end{align}
for the second term in brackets in \myEqua{eq:BaxialEQM}. 
It contains a constant component and an oscillatory term at the difference frequency~$\myombw= \myompw-\myommw$ of the two radial modes, but resonant terms at the axial frequency~$\myomzw$ naturally result from the axial motion. 
Expressed in our notation, the mechanism reads
\begin{gather}
\myfComp{\myZOT{z}^{{\myinn}-2k+1}\myZOT{\rho}^{2k-2}\myZOT{\xi}}{\myomzw}= \myfComp{\myZOT{z}^{{\myinn}-2k+1}}{\myomzw} \myfComp{\myZOT{\rho}^{2k-2}\myZOT{\xi}}{\myfZero} \myGSAbstand. 
\end{gather}
According to \myEquas{eq:cos2N} and~\eqref{eq:cos2Np1}, the exponent~$\myinn-2k+1$ has to be odd for $\myZOT{z}^{\myinn-2k+1}$ to have a term at the fundamental frequency~$\myomzw$. 
Hence, $\myinn$ has to be even. 
We incorporate this parity by letting $\myinn=2n$.
The resonant terms which arise from \myEqua{eq:BaxialEQM} are then given by
\begin{align}
\gamma_z \myZOT{z} 
&= - \frac{\myomc}{\myomz^2} \frac{B_{2n}}{B_0}
\sum_{k=1}^{n} \myatnk{2n}{k}\, \myfComp{\myZOT{z}^{2(n-k)+1}}{\myomzw} \myfComp{\myZOT{\rho}^{2(k-1)}\myZOT{\xi}}{\myfZero} \label{eq:Bn-axial-gammazZOT} \myGSAbstand.
\end{align}
These terms are to be absorbed in the effective zeroth-order equation of motion~\eqref{eq:AxialEQM-Eff}. 
The contribution by the axial oscillation is read off from \myEqua{eq:cos2Np1FundFreq}, and the coefficient~$\myatnk{2n}{k}$ is defined in \myEqua{eq:myantk}. 
The non-oscillatory contribution of the radial modes results from two terms summarized as 
\begin{gather}
\myfComp{\myZOT{\rho}^{2(k-1)} \myZOT{\xi}}{\myfZero} = \myfComp{\myZOT{\rho}^{2(k-1)}}{\myfZero} \myfComp{\myZOT{\xi}}{\myfZero}
 + \myfComp{\myfComp{\myZOT{\rho}^{2(k-1)}}{\myombw} \myfComp{\myZOT{\xi}}{\myombw}}{\myfZero} \label{eq:Bn-Axial-RadZero}\myGSAbstand.
\end{gather}
Clearly, $\myZOT{\rho}^{2(k-1)}$ and $\myZOT{\xi}$ possess a constant component. 
However, mixing the oscillatory contributions at $\myombw$ yields a constant component, too. 
With \myEquas{eq:rhoqN-DC} and~\eqref{eq:xiZOT}, we have
\begin{align}
\begin{split}
\myfComp{\myZOT{\rho}^{2(k-1)}}{\myfZero} \myfComp{\myZOT{\xi}}{\myfZero} &= \myompw\myAmpR{+}^2\left\{\sum_{p=0}^{k-1} \left[\binom{k-1}{p}\right]^2 \myAmpR{+}^{2p} \myAmpR{-}^{2(k-1-p)}\right\} \\& \quad + \myommw\myAmpR{-}^2\left\{\sum_{p=0}^{k-1} \left[\binom{k-1}{p}\right]^2 \myAmpR{-}^{2p} \myAmpR{+}^{2(k-1-p)}\right\} \label{eq:Baxial-RadZero}
\end{split}
\end{align}
for the first term. 
We have used the freedom in the choice of the two radial amplitudes in \myEqua{eq:rhoqN-DC} such that the two sums in \myEqua{eq:Baxial-RadZero} have the same structure for both amplitudes. 
We will do the same with the choices in \myEqua{eq:rhoqN-FF} and we obtain
\begin{multline}
\myfComp{\myfComp{\myZOT{\rho}^{2(k-1)}}{\myombw} \myfComp{\myZOT{\xi}}{\myombw}}{\myfZero} 
= \myompw\myAmpR{+}^2\left\{\sum_{p=0}^{k-1}\binom{k-1}{p}\binom{k-1}{p+1} \myAmpR{+}^{2p}\myAmpR{-}^{2(k-1-p)}\right\} \\
  + \myommw\myAmpR{-}^2\left\{\sum_{p=0}^{k-1}\binom{k-1}{p}\binom{k-1}{p+1}\myAmpR{-}^{2p}\myAmpR{+}^{2(k-1-p)}\right\}
\label{eq:Baxial-RadZeroBB}
\end{multline}
for the second term in \myEqua{eq:Bn-Axial-RadZero}. 
\myEquas{eq:Baxial-RadZero} and~\eqref{eq:Baxial-RadZeroBB} share one common binomial coefficient. 
The other two are summed with the identity~\eqref{eq:BinoAddId1} in order to give
\begin{multline}
\myfComp{\myZOT{\rho}^{2(k-1)} \myZOT{\xi}}{\myfZero} =  \myompw\myAmpR{+}^2\left\{\sum_{p=0}^{k-1}\binom{k-1}{p}\binom{k}{p+1} \myAmpR{+}^{2p}\myAmpR{-}^{2(k-1-p)}\right\} \\
  + \myommw\myAmpR{-}^2\left\{\sum_{p=0}^{k-1}\binom{k-1}{p}\binom{k}{p+1}\myAmpR{-}^{2p}\myAmpR{+}^{2(k-1-p)}\right\} \myGSAbstand.
\end{multline}
Finally, the two sums with a prefactor of $\myompw$ and $\myommw$ are combined by transforming the summation variable as $p \to p-1$ in
\begin{gather}
\myAmpR{+}^2\sum_{p=0}^{k-1}\binom{k-1}{p}\binom{k}{p+1} \myAmpR{+}^{2p}\myAmpR{-}^{2(k-1-p)}= \sum_{p=0}^{k}\binom{k-1}{p-1}\binom{k}{p} \myAmpR{+}^{2p}\myAmpR{-}^{2(k-p)}
\end{gather}
and as $p\to k-p-1$ in
\begin{align}
\myAmpR{-}^2 \sum_{p=0}^{k-1}  \binom{k-1}{p}\binom{k}{p+1} \myAmpR{-}^{2p} \myAmpR{+}^{2(k-1-p)}  
 &=\sum_{p=0}^{k}  \binom{k-1}{p}\binom{k}{p} \myAmpR{+}^{2p} \myAmpR{-}^{2(k-p)} \myGSAbstand, 
\end{align}
which results in the same exponent $2p$ for $\myAmpR{+}$, and $2(k-p)$ for $\myAmpR{-}$ in both sums. 
Note that we have matched the limits of the two sums by including vanishing contributions at the lower and the upper limit, respectively. 
The non-oscillatory contribution from the radial modes is then found as
\begin{align}
\myfComp{\myZOT{\rho}^{2(k-1)} \myZOT{\xi}}{\myfZero}& =\sum_{p=0}^{k} \binom{k}{p}\left[\myompw\binom{k-1}{p-1} + \myommw\binom{k-1}{p}\right] \myAmpR{+}^{2p} \myAmpR{-}^{2(k-p)} \label{eq:B-axial-rhoXi-Zero-total} \myGSAbstand.
\end{align}

All the contributions to the parameter~$\gamma_z$ from \myEqua{eq:Bn-axial-gammazZOT} have now been calculated, but before relating~$\gamma_z$ to the first-order shift of the axial frequency via \myEqua{eq:DeltaOmz}, we note that \myEqua{eq:B-axial-rhoXi-Zero-total} still contains the perturbed radial frequencies~$\myompmw$. 
Since the difference between $\myompmw$ and the radial frequencies $\myompm$ in the ideal trap is at least of first order in a perturbation parameter (not necessarily $B_{2n}$ alone), and $\gamma_z$ is of first order in $B_{2n}$, we can replace  $\myompmw$ with the unperturbed frequencies $\myompm$ without incurring an error of first order in the frequency-shift.
 
The fact that the first-order frequency-shift is by definition of first order in a perturbation parameter also has a welcome practical consequence. 
For calculating first-order frequency shifts, it does not matter whether the perturbed and presumably measured frequency~$\myomw_i$ or the frequency~$\omega_i$ that the particle would have in the limit of zero imperfections is used. 
The overall difference in the frequency-shift is of second order, which includes cross-terms between different perturbation parameters. 
For our case of cylindrically-symmetric imperfections, the limit of no imperfections is attained for vanishing amplitudes as far as quantum mechanics permits, but the effect of other imperfections may be harder to measure and to correct for. 
In total, the ``true'' frequency~$\omega_i$ the particle would have in the fully classical ideal Penning trap may remain unknown until corrections are applied.

All things considered, the first-order shift to the axial frequency is given  by 
\begin{align}
\begin{split}
\frac{\Delta \myomz}{\myomz}  &= -\frac{B_{2n}}{B_0}\frac{\myomc}{\myomp\myomm} \frac{(2n)!}{2^{2n+1}}\sum_{k=1}^{n} \frac{(-1)^k\,k}{[k!(n-k)!]^2}\frac{\myAmpZ^{2(n-k)}}{n-k+1}
\\& \quad \cdot \sum_{p=0}^{k} \binom{k}{p}\left[\myomp\binom{k-1}{p-1} + \myomm\binom{k-1}{p}\right] \myAmpR{+}^{2p} \myAmpR{-}^{2(k-p)} \myGSAbstand,
\end{split} \label{eq:Bn-FS-Axial} 
\end{align}
where we have used $\myomz^2=2\myomp\myomm$ from \myEqua{eq:omzq-omrad} in order to write the right-hand side with only the frequencies of the radial modes. 
We refrain from expressing the binomial coefficients explicitly with factorials because we would have to deal with two exceptions. 
The term with $\myomp$ does not contribute for $p=0$; 
the term with $\myomm$ vanishes for $p=k$. 
The definition of the binomial coefficient in \myEqua{eq:BinoDef} has these exceptions covered more conveniently than splitting the sum or introducing Kronecker deltas. 

Since a typical experiment has $\lvert \myomp \rvert \gg \lvert\myomm\rvert$, one might be inclined to neglect the term with $\myomm$, but some care has to be taken to avoid losing a degree of freedom. 
As we have seen, the term with $\myomm$ is the only contribution for $p=0$. 
Moreover, the scaling of the two binomial coefficients is such that the term with $\myomm$ is boosted compared to the other term for the combination of small $p$ and large $k$, and hence large $n$. 
Since we cannot give an approximation that is valid for all $n$, we stick with the full expression, leaving it to the reader to neglect certain terms after having evaluated the formula for a specific~$n$ and a set of radial frequencies.

Unlike the axial-frequency shift caused by electrostatic imperfections, the shift described by \myEqua{eq:Bn-FS-Axial} is not symmetric with respect to the amplitudes of the two radial modes. 
Technically, this imbalance is a consequence of \myEqua{eq:xiZOT}, where the $\myAmpR{\pm}^2$ enter with their respective frequency~$\myompmw$. 
On more physical grounds, the force caused by the additional magnetic field depends on velocities. 
For the same radial amplitudes, the velocity of the modified cyclotron motion is larger than the one of the magnetron motion by a factor of $\myomp/\myomm$.

The axial-frequency shift caused by $B_2$ is often estimated by assigning a magnetic moment to the radial eigenmotions as the orbiting charge can be considered a circular current. 
The shift then results from the coupling of the averaged magnetic moment of the radial motions to the axial magnetic field. 
This intuitive model works fine for $B_2$, but it obscures the more general mechanism. 
It is the radial magnetic field that couples the radial motion to the axial mode. 
However, the axial and the radial magnetic field are related since they originate from the same scalar potential defined in \myEqua{eq:Psi-rnPn}.

Unfortunately, a direct crosscheck of our result in \myEqua{eq:Bn-FS-Axial} with the equally general Equation~(3.73) in~\cite{cptI2005} is of no avail because the latter must be dismissed on a simple dimensional argument. 
%
\subsection{Radial modes}
The radial equations of motion are the first two components in \myEqua{eq:RealBEqMotions}.
Unlike for the axial mode, which contains only the radial components of the additional magnetic field, the axial component shows up here, too. 
We will first examine which of the two components leads to natural coherence. 
Recalling the radial magnetic field given in \myEqua{eq:Bn-radCart}, we note that the terms from the axial mode are of the kind 
\begin{gather}
\dot{z}z^{\myinn-2k+1} = \frac{1}{\myinn-2k+2} \frac{\myd}{\myd t}z^{\myinn-2k+2} \myGSAbstand,
\end{gather}
where we have used the time-derivative in the last step. 
By inserting the zeroth-order solution~$\myZOT{z}(t)$, we will get a sum of single-frequency oscillatory terms. 
The discrete frequencies in the frequency-spectrum of $\myZOT{z}^{\myinn-2k+2}$ are not changed by taking the time-derivative; 
only the weights are affected. 
Most notably, a non-oscillatory component is removed by the time-derivative. 
Therefore, there is no constant term in $\dot{\myZOT{z}}\myZOT{z}^{\myinn-2k+1}$. 
Resonant terms at the radial frequencies naturally arise from the radial modes, but the initial oscillatory terms at the right frequency in $B_{\myinn}^{(x)}$ and  $B_{\myinn}^{(y)}$ are rendered off-resonant through mixing with the axial frequency or its higher harmonics. 
Consequently, the radial components of the additional magnetic field do not give rise to a first-order frequency-shift and they are ignored for the remainder of the calculation. 
The radial equations of motion then are effectively simplified to 
\begin{align}
\begin{pmatrix}\ddot{x}\\\ddot{y}\end{pmatrix} 
&=\myomc \begin{pmatrix}\dot{y}\\-\dot{x}\end{pmatrix}
\left[1 + \frac{B_{\myinn}}{B_0}\sum_{k=0}^{\left\lfloor \myinn/2 \right\rfloor} \myank{\myinn}{k}\; z^{\myinn-2k} \rho^{2k}
\right]
 + \frac{\myomz^2}{2}\begin{pmatrix}x\\y\end{pmatrix} 
\myGSAbstand,
\end{align}
where we have expressed the additional axial magnetic field according to \myEqua{eq:BnAxial}. 
We perform the frequency-analysis as before by plugging in the zeroth-order solutions from \myEquas{eq:z-ZOT} and~\eqref{eq:xyZOT}.

The mechanism for producing resonant terms is completely analogous to the treatment of the radial modes in \myEqua{eq:CnRadEQM} for electrostatic imperfections and it reads
\begin{gather}
\myfComp{\myZOT{z}^{\myinn-2k}\myZOT{\rho}^{2k} \dot{\myZOT{y}}}{\myompmw} = \myfComp{\myZOT{z}^{\myinn-2k}}{\myfZero}\myfComp{\myZOT{\rho}^{2k} \dot{\myZOT{y}}}{\myompmw} 
\end{gather}
in our notation. 
Everything else carries over from the first treatment of the radial modes in \mySubsec{subsec:CRadModes}, too. 
There is a constant term in $\myZOT{z}^{\myinn-2k}$ only for $\myinn-2k$ even, and hence we have to choose $\myinn$ even. 
With the definition~$\myinn=2n$, the resonant terms in the $x$-component of the radial equations of motion are written as 
\begin{align}
\beta_\pm\dot{\myZOT{y}}_\pm 
&= \frac{B_{2n}}{B_0}\sum_{k=0}^{n} \myank{2n}{k}\, \myfComp{\myZOT{z}^{2(n-k)}}{\myfZero} \myfComp{\myZOT{\rho}^{2k} \dot{\myZOT{y}}}{\myompmw} \label{eq:betapm-yv} \myGSAbstand.
\end{align}
Our goal is to end up with the effective equation of motion~\eqref{eq:radEQM-Eff}. 
The constant component from the axial oscillation is calculated with \myEqua{eq:cos2NConst}; 
the coefficient $\myank{2n}{k}$ is defined in \myEqua{eq:ank}.  
The velocity $\dot{\myZOT{y}}$ is found by taking the time-derivative of \myEqua{eq:xyZOT} and just like the coordinate~$\myZOT{y}(t)$ it contains a contribution $\dot{\myZOT{y}}_\pm(t)$ at the frequency~$\myompmw$. 
Thus, $\dot{\myZOT{y}}_\pm(t)$ stays resonant right away if it is multiplied with constant terms only, and $\dot{\myZOT{y}}_\pm(t)$ becomes resonant at the other radial frequency~$\myommpw$ if it is mixed with a term at the frequency~$\myombw=\myompw-\myommw$. 
Naturally, such a term results from powers of the radial displacement squared. 
Thus, oscillatory terms at the radial frequencies are produced by the two mechanism formally expressed as
\begin{align}
\myfComp{\myZOT{\rho}^{2k} \dot{\myZOT{y}}}{\myompmw}&= \myfComp{\myZOT{\rho}^{2k}}{\myfZero}\dot{\myZOT{y}}_{\pm} + \myfComp{\myfComp{\myZOT{\rho}^{2k}}{\myombw} \dot{\myZOT{y}}_{\mp}}{\myompmw}\myGSAbstand.
\end{align}
The first term on the right-hand side is calculated with the help of \myEqua{eq:rhoqN-DC}.
Combining \myEquas{eq:rhoqN-FF} and~\eqref{eq:cosBeatcospm} yields
\begin{align}
\myfComp{ \myfComp{\myZOT{\rho}^{2k}}{\myombw} \dot{\myZOT{y}}_{\mp}}{\myompmw} 
&= \dot{\myZOT{y}}_\pm\, \frac{\myommpw}{\myompmw}\sum_{p=0}^{k}\binom{k}{p}\binom{k}{p+1}\, \myAmpR{\pm}^{2p}\myAmpR{\mp}^{2(k-p)} 
\end{align}
for the second term. 
We have chosen the combination~$\myAmpR{\pm}$ in \myEqua{eq:rhoqN-FF} such that the factor $\myAmpR{\pm}/\myAmpR{\mp}$ compensates for the factor $\myAmpR{\mp}/\myAmpR{\pm}$ that results from producing $\dot{\myZOT{y}}_\pm$ out of $\dot{\myZOT{y}}_\mp$ through mixing. 
Recall that $\dot{\myZOT{y}}_\mp$ comes with $\myommpw\myAmpR{\mp}$, whereas an additional factor of $\myompmw\myAmpR{\pm}$ is necessary to create $\dot{\myZOT{y}}_\pm$ after mixing with the oscillatory component at $\myombw$.
In total, the resonant contribution from the radial modes is 
\begin{align}
\myfComp{\myZOT{\rho}^{2k}\dot{\myZOT{y}}}{\myompmw}
&= \dot{\myZOT{y}}_\pm \sum_{p=0}^{k} \binom{k}{p} \left[\binom{k}{p} + \frac{\myommpw}{\myompmw}\binom{k}{p+1}\right]\, \myAmpR{\pm}^{2p}\myAmpR{\mp}^{2(k-p)} \myGSAbstand.
\end{align}
All terms considered, the parameter~$\beta_\pm$ in \myEqua{eq:betapm-yv} is determined as
\begin{gather}
\begin{split}
\beta_\pm &= \frac{B_{2n}}{B_0}\frac{(2n)!}{2^{2n}} \sum_{k=0}^{n}\frac{(-1)^{k}\,  \myAmpZ^{2(n-k)}}{[k!(n-k)!]^2}  \\ &\quad\cdot \sum_{p=0}^{k} \binom{k}{p} \left[\binom{k}{p} + \frac{\myommpw}{\myompmw}\binom{k}{p+1}\right] \myAmpR{\pm}^{2p}\myAmpR{\mp}^{2(k-p)} 
\end{split}
\end{gather}
and related to the first-order frequency-shift 
\begin{align}
\begin{split}
\Delta \myompm&= \pm\frac{B_{2n}}{B_0}\frac{\myomc}{\myomp - \myomm}\frac{(2n)!}{2^{2n}} \sum_{k=0}^{n}\frac{(-1)^{k}\, \myAmpZ^{2(n-k)} }{[k!(n-k)!]^2}  \\ &\quad\cdot\sum_{p=0}^{k} \binom{k}{p} \left[\myompm\binom{k}{p} + \myommp\binom{k}{p+1}\right] \myAmpR{\pm}^{2p}\myAmpR{\mp}^{2(k-p)} \label{eq:Bn-FS-Radial} 
\end{split}
\end{align}
via \myEqua{eq:RadFoFs}. 
Along the lines described for $\gamma_z$ and the shifts caused by $B_{2n}$ to the axial frequency in \mySubsec{subsec:MagAxialMode}, we have replaced the perturbed frequencies~$\myompmw$ in $\beta_\pm$ with $\myompm$. 
The error incurred by this substitution is at least of second order in the frequency-shift and hence does not affect the first-order result.

By sending the summation variable $p\to k-p$, we obtain
\begin{gather}
\begin{split}
\Delta \myompm &= \pm\frac{B_{2n}}{B_0}\frac{\myomc}{\myomp - \myomm}\frac{(2n)!}{2^{2n}} \sum_{k=0}^{n}\frac{(-1)^{k}\, \myAmpZ^{2(n-k)}}{[k!(n-k)!]^2}  \\ &\quad\cdot \sum_{p=0}^{k} \binom{k}{p} \left[\myompm\binom{k}{p} + \myommp\binom{k}{p-1}\right] \myAmpR{\mp}^{2p}\myAmpR{\pm}^{2(k-p)} \label{eq:Bn-FS-Radial2}
\end{split}
\end{gather}
and we confirm that the frequency-shift is not symmetric with respect to the two amplitudes of the radial modes. 
Given the mechanism of producing an oscillatory term at~$\myompmw$ by mixing an oscillatory term at $\myommpw$ with the symmetric contribution at $\myombw$ from powers of radial ion displacement squared, the imbalance comes as no surprise. 

Note that the binomial coefficient with the factor of $\myommp$ does not contribute for $p=k$ in \myEqua{eq:Bn-FS-Radial}, and for $p=0$ in \myEqua{eq:Bn-FS-Radial2}. 
Because of this exception, we will not use the explicit expression from \myEqua{eq:BinoDef} for the binomial coefficient. 
The exception should also be taken into account when neglecting terms with a factor of $\myomp$ against those with a factor of $\myomm$. 
Additionally, the scaling of the two relevant binomial coefficients is such that the factor of $\myomp/\myomm$ can be outweighed for large $k$, and hence large $n$. 
Just like for the axial-frequency shift caused by magnetic imperfections, there is no generally valid approximation for all~$n$ here, even for $\lvert \myomp\rvert \gg \lvert \myomm\rvert$.

A direct comparison of our result in \myEqua{eq:Bn-FS-Radial} with Equation~(3.74) in~\cite{cptI2005} is inconclusive because the latter must again be ruled out on a simple dimensional argument. 

\paragraph{The cyclotron sideband}
By adding \myEquas{eq:Bn-FS-Radial} and~\eqref{eq:Bn-FS-Radial2} for the two different radial modes, the shift to the sideband cyclotron-frequency is expressed as
\begin{align}
\begin{split}
\Delta \myomc  &= 
\frac{B_{2n}}{B_0}\frac{\myomc}{\myomp - \myomm}\frac{(2n)!}{2^{2n}} \sum_{k=0}^{n}\frac{(-1)^{k} \,\myAmpZ^{2(n-k)}}{[k!(n-k)!]^2}   \sum_{p=0}^{k} \binom{k}{p} \\ &\quad\cdot\left\{\myomp\left[\binom{k}{p} - \binom{k}{p-1}\right]+\myomm\left[\binom{k}{p+1}-\binom{k}{p}\right]\right\}  \myAmpR{+}^{2p}\myAmpR{-}^{2(k-p)} \myGSAbstand.
\end{split} \label{eq:Bn-FS-SB}
\end{align}
%
\section{Explicit expressions for frequency-shifts}
Having derived the general formulas in \mySecs{sec:FreqShiftC} and \ref{sec:FreqShiftB}, we generate explicit expression for the most frequently-used lowest-order imperfections, thereby also demonstrating that we have the most comprehensive specific treatment~\cite{rainville2003} covered. 
For the three lowest-order electrostatic imperfections, we evaluate the first-order frequency-shifts as given by \myEquas{eq:Cn-FS-Axial}, \eqref{eq:Cn-FS-Radial} and~\eqref{eq:Cn-FS-SB} for the axial frequency~$\myomz$, the two radial frequencies~$\myompm$ and the sideband cyclotron-frequency, respectively.
The shifts caused by magnetic imperfections to the aforementioned frequencies are given in \myEquas{eq:Bn-FS-Axial}, \eqref{eq:Bn-FS-Radial} and~\eqref{eq:Bn-FS-SB}, and we evaluate these formulas for the two lowest-order magnetic imperfections. 
The frequency-shift is defined as the difference between the perturbed and the unperturbed frequency. 
The axial frequency in the ideal Penning trap is given in \myEqua{eq:uptFreqAxial}; 
the two frequencies~$\myompm$ of the radial modes are defined in \myEqua{eq:uptRadFreq}. 
The quadrupole potential of the ideal trap is given in \myEqua{eq:IPT-quadPot}, and the ideal magnetic field is $\myvec{B}_0 = B_0\myeuv{z}$. 

To first-order in the frequency-shift, the parameters $\myAmpR{+}$, $\myAmpR{-}$ and $\myAmpZ$ can be identified as the amplitudes of the modified cyclotron, magnetron and axial motion, respectively. 
We stress that the first-order frequency-shifts caused by multiple imperfections add up linearly. 
Moreover, the imperfections associated with $\myinn$ odd do not give rise to a first-order frequency-shift.

For reference and better comparison with other definitions of the perturbation parameters~$C_{\myinn}$ and $B_{\myinn}$, we also show the corresponding higher-order electrostatic potential (according to \myEquas{eq:PhiN-rTheta} and~\eqref{eq:rnPn-zRho}) and magnetic field (calculated from \myEquas{eq:BnAxial} and~\eqref{eq:Bn-rad}), respectively.

\subsection{Electrostatic imperfections}
%
\begin{gather}
\Phi_4 = C_4\frac{V_0}{2d^4}\left(z^4-3 z^2 \rho^2+\frac{3}{8} \rho^4\right)
\end{gather}
\begin{align}
\frac{\Delta \myomz}{\myomz} &= \frac{C_4}{C_2}\frac{3}{4d^2} \left(\myAmpZ^2 - 2\myAmpR{+}^2 - 2\myAmpR{-}^2\right) \\
\Delta \myompm &=\mp\frac{C_4}{C_2} \frac{3}{2d^2}\frac{\myomp\myomm}{\myomp-\myomm}\left(2\myAmpZ^2 - \myAmpR{\pm}^2-2\myAmpR{\mp}^2\right) \\
\Delta \myomc&= -\frac{C_4}{C_2}\frac{3}{2d^2}\frac{\myomp\myomm}{\myomp-\myomm}\left(\myAmpR{+}^2-\myAmpR{-}^2\right)
\end{align}
\begin{gather}
\Phi_6 = C_6\frac{V_0}{2d^6}\left(z^6-\frac{15}{2} z^4 \rho^2+\frac{45}{8} z^2 \rho^4-\frac{5}{16} \rho^6 \right)
\end{gather}
\begin{align}
\begin{split}
\frac{\Delta \myomz}{\myomz} &= \frac{C_6}{C_2}\frac{15}{16d^4} \left(\myAmpZ^4 +3\myAmpR{+}^4+ 3\myAmpR{-}^4 \right.\\ &\left. \quad {} - 6\myAmpR{+}^2\myAmpZ^2 - 6\myAmpR{-}^2\myAmpZ^2 + 12\myAmpR{+}^2\myAmpR{-}^2\right) 
\end{split} \\
\begin{split}
\Delta \myompm &= \mp\frac{C_6}{C_2} \frac{15}{8d^4}\frac{\myomp\myomm}{\myomp-\myomm} \left(3\myAmpZ^4 + \myAmpR{\pm}^4+3\myAmpR{\mp}^4 -6\myAmpR{\pm}^2\myAmpZ^2\right. \\ &\quad \left. {} -12\myAmpR{\mp}^2\myAmpZ^2+6\myAmpR{+}^2 \myAmpR{-}^2\right) 
\end{split} \\
\Delta \myomc 
&= \frac{C_6}{C_2} \frac{15}{4d^4} \frac{\myomp\myomm}{\myomp-\myomm} \left( \myAmpR{+}^2-\myAmpR{-}^2 \right) \left(-3\myAmpZ^2 +\myAmpR{+}^2 +\myAmpR{-}^2 \right)
\end{align}
\begin{gather}
\Phi_8 = C_8\frac{V_0}{2d^8}\left(z^8- 14 z^6 \rho^2 + \frac{105}{4} z^4 \rho^4-\frac{35}{4} z^2 \rho^6 + \frac{35}{128}\rho^8\right)
\end{gather}
\begin{align}
\begin{split}
\frac{\Delta \myomz}{\myomz} &= \frac{C_8}{C_2}\frac{35}{32d^6}\left(\myAmpZ^6 -4\myAmpR{+}^6 - 4\myAmpR{-}^6 +18\myAmpR{+}^4\myAmpZ^2+18\myAmpR{-}^4\myAmpZ^2\right.\\
&\left.\quad {} - 36\myAmpR{+}^4\myAmpR{-}^2 - 36\myAmpR{+}^2\myAmpR{-}^4 +72\myAmpR{+}^2\myAmpR{-}^2\myAmpZ^2 - 12\myAmpR{+}^2\myAmpZ^4 - 12\myAmpR{-}^2\myAmpZ^4\right) 
\end{split}\\
\begin{split}
\Delta \myompm &=\mp\frac{C_8}{C_2} \frac{35}{16d^6}\frac{\myomp\myomm}{\myomp-\myomm}\left(4\myAmpZ^6-\myAmpR{\pm}^6 - 4\myAmpR{\mp}^6 + 12\myAmpR{\pm}^4\myAmpZ^2 +36\myAmpR{\mp}^4\myAmpZ^2 \right. \\
& \quad\left. {} -12\myAmpR{\pm}^4\myAmpR{\mp}^2-18\myAmpR{\pm}^2\myAmpR{\mp}^4 - 18\myAmpR{\pm}^2\myAmpZ^4 - 36\myAmpR{\mp}^2\myAmpZ^4 +72\myAmpR{+}^2\myAmpR{-}^2\myAmpZ^2\right) 
\end{split}\\
\begin{split}
\Delta \myomc &=-\frac{C_8}{C_2}\frac{105}{16d^6}\frac{\myomp\myomm}{\myomp-\myomm}  \left( \myAmpR{+}^2-\myAmpR{-}^2 \right)\left(6\myAmpZ^4 + \myAmpR{+}^4 + \myAmpR{-}^4 \right. \\
&\quad  \left. {} - 8\myAmpR{+}^2\myAmpZ^2 - 8 \myAmpR{-}^2\myAmpZ^2 +3\myAmpR{+}^2\myAmpR{-}^2\right) 
\end{split}
\end{align}
%
\subsection{Magnetostatic imperfections}
%
\begin{gather}
\myvec{B}_2 = B_2\left[\left(z^2-\frac{1}{2}\rho^2 \right)\myeuv{z} + \left(-z\rho \right)\myeuv{\rho}\right]
\end{gather}
\begin{align}
\frac{\Delta \myomz}{\myomz} &= \frac{B_2}{4B_0}\frac{\myomp+\myomm}{\myomp\myomm} \left(\myAmpR{-}^2 \myomm+\myAmpR{+}^2 \myomp\right)\\
\frac{\Delta \myomp}{\myomp} &= \frac{B_2}{2B_0}\frac{\myomp+\myomm}{\myomp-\myomm} \left[\myAmpZ^2 - \myAmpR{+}^2 -\myAmpR{-}^2 \left( 1 + \frac{\myomm}{\myomp} \right) \right]  \\
\frac{\Delta \myomm}{\myomm} &= -\frac{B_2}{2B_0}\frac{\myomp+\myomm}{\myomp-\myomm} \left[\myAmpZ^2  - \myAmpR{+}^2 \left( \frac{\myomp}{\myomm} + 1 \right) - \myAmpR{-}^2\right]  \\
\frac{\Delta \myomc}{\myomc}&= \frac{B_2}{2B_0} \left[\myAmpZ^2- \myAmpR{-}^2\frac{\myomp}{\myomp-\myomm} +\myAmpR{+}^2\frac{\myomm}{\myomp-\myomm}\right] 
\end{align}
\begin{gather}
\myvec{B}_4 = B_4\left[\left(z^4-3 z^2 \rho^2+\frac{3}{8} \rho^4\right)\myeuv{z} + \left(-2 z^3 \rho +\frac{3}{2} z \rho^3 \right)\myeuv{\rho}\right]
\end{gather}
\begin{align}
\begin{split}
\frac{\Delta \myomz}{\myomz} &= \frac{3B_4}{8B_0}\frac{\myomp+\myomm}{\myomp\myomm}\left[\myomm\left(-\myAmpR{-}^4 +\myAmpR{-}^2\myAmpZ^2 -2\myAmpR{+}^2\myAmpR{-}^2\right)\right.\\ &\quad\left. {} + \myomp\left(-\myAmpR{+}^4+ \myAmpR{+}^2\myAmpZ^2- 2\myAmpR{+}^2 \myAmpR{-}^2 \right)\right]
\end{split}
\end{align}
\begin{align}
\begin{split}
\frac{\Delta \myomp}{\myomp} &= \frac{3B_4}{8B_0}\frac{\myomp+\myomm}{\myomp-\myomm} \left[\myAmpZ^4  + \myAmpR{+}^4 + \myAmpR{-}^4 \left(1+2\frac{\myomm}{\myomp} \right) -4\myAmpR{+}^2\myAmpZ^2 \right. \\
&\quad \left. {} - 4\myAmpR{-}^2\myAmpZ^2\left(1+\frac{\myomm}{\myomp}\right) + 4\myAmpR{+}^2\myAmpR{-}^2\left(1+\frac{\myomm}{2\myomp}\right)\right] 
\end{split} \\
\begin{split}
\frac{\Delta \myomm}{\myomm} &= -\frac{3B_4}{8B_0} \frac{\myomp+\myomm}{\myomp-\myomm} \left[  \myAmpR{+}^4\left(2\frac{\myomp}{\myomm}+1\right) + 2\myAmpR{+}^2\myAmpR{-}^2\left(\frac{\myomp}{\myomm}+2\right)  \right.\\
& \quad \left. {} - 4\myAmpR{+}^2\myAmpZ^2\left(\frac{\myomp}{\myomm}+1\right)+ \myAmpZ^4  + \myAmpR{-}^4 - 4\myAmpR{-}^2\myAmpZ^2 \right] 
\end{split} \\
\begin{split}
\frac{\Delta \myomc}{\myomc}&= \frac{3B_4}{8B_0}\left[\vphantom{\frac{1}{1}}\myAmpZ^4  + 2 \myAmpR{+}^2 \myAmpR{-}^2 \right. \\ &\quad \left. {} + \frac{\myomc\left(-\myAmpR{+}^4+\myAmpR{-}^4\right) + 4 \myAmpZ^2 \left(\myAmpR{+}^2\myomm - \myAmpR{-}^2\myomp\right) }{\myomp - \myomm}  \right] 
\end{split}
\end{align}
%
\section{Conclusion}
By identifying the mechanism for producing terms that are in phase with the motions at the fundamental eigenfrequencies, we have calculated the first-order frequency-shifts caused by static cylindrically-symmetric electric and magnetic imperfections of a Penning trap consistently for all perturbation parameters~$C_\myinn$ and $B_\myinn$, culminating in general expressions for the shifts to all three eigenfrequencies. 
The easy evaluation of the fully analytic expression is enabled by  a general parametrization of the imperfections in cylindrical instead of spherical coordinates. 
The explicit link between $r^{\myinn} P_{\myinn}(\cos(\theta))$ and cylindrical coordinates has often been missed by Penning-trap literature, performing the transformation of coordinates separately for each~$\myinn$.

\section*{Acknowledgments}
We thank Sven Sturm for helpful comments. 
This work was funded by the Max-Planck-Gesellschaft and the ERC Grant Precision Measurements of Fundamental Constants (MEFUCO). 
T.\,E.\ was supported by a fellowship of the Alexander von Humboldt foundation. 
S.\,S. acknowledges support by the Heidelberg Graduate School of Fundamental Physics (HGSFP). 
J.\,K.\ acknowledges support by the HGSFP and by the International Max Planck Research School for Precision Tests of Fundamental Symmetries (IMPRS-PTFS). 



\end{document}